\documentstyle[preprint,aps] {revtex}                 %%%%%  new REVTeX 3.0
\begin{document}
%\draft
\title{Determination of the $E2/M1$ Ratio in the $\gamma N \rightarrow \Delta(1232)$ Transition 
from a Simultaneous Measurement of $p(\vec{\gamma},p)\pi^0$ and $p(\vec{\gamma},\pi^+)n$ }

\author{ R.~Beck$^1$, H.P.~Krahn$^1$, J.~Ahrens$^1$, J.R.M.~Annand$^5$,
H.J.~Arends$^1$, G. Audit$^2$, A.~Braghieri$^3$, N.~d'Hose$^2$,
D.~Drechsel$^1$, O.~Hanstein$^1$, J.C.~McGeorge$^5$, R.O.~Owens$^5$,
P.~Pedroni$^3$, T.~Pinelli$^{3,4}$, G.~Tamas$^2$, L.~Tiator$^1$ and
Th.~Walcher$^1$}
\address{$^1$Institut f\"ur Kernphysik, Universit\"at Mainz, 55099 Mainz, Germany\\
$^2$Service de Physique Nucl\'{e}aire--DAPNIA, CEA. Saclay, 91191 Gif--sur--Yvette, France\\
$^3$Istituto Nazionale di Fisica Nucleare, Sezione di Pavia, 27100 Pavia, Italy\\
$^4$Dipartimento di Fisica Nucleare e Teorica, Universit\`{a} di Pavia, 27100 Pavia, Italy\\
$^5$Department of Physics and Astronomy, Glasgow University, Glasgow G12 8QQ, UK}

\date{\today}

\maketitle

\begin{abstract}
Tagged linearly polarized photons have been used at the Mainz Microtron 
MAMI for simultaneous measurements of the $p(\vec{\gamma},p)\pi^0$ and
$p(\vec{\gamma},\pi^+)n$ reaction channels to study the 
$\gamma N \rightarrow \Delta(1232)$ transition. The energy dependence of 
the magnetic dipole 
$M_{1+}^{3/2}$ and electric quadrupole $E_{1+}^{3/2}$ amplitudes have been 
extracted from these data in the photon energy range from $270$ to 
$420 {\rm MeV}$. The $E2/M1$ ratio for the 
$\gamma N \rightarrow \Delta(1232)$ transition has been determined to 
be $- (2.5\pm0.1_{stat}\pm0.2_{sys})\%$ at the resonance position 
($\delta_{33}=90^0$).\\
\end{abstract}

%
%\begin{enumerate}
%

\section{\bf Introduction}
%
%\twocolumn
%

Low energy electromagnetic properties of baryons, such as mass, charge
radius, magnetic and quadrupole moments are important observables for any
model of the nucleon structure. In various constituent--quark models
a tensor force in the inter--quark hyperfine interaction, introduced first
by de Rujula, Georgi and Glashow \cite{Ruj75}, leads to a $d$--state 
admixture in the baryon ground state wave function. As a result the tensor 
force induces a small violation of the Becchi--Morpurgo selection rule 
\cite{Bec65}, that the $\gamma N \rightarrow \Delta(1232)$ excitation
is a pure $M1$ (magnetic dipole) transition, by introducing a 
non--vanishing $E2$ (electric quadrupole) amplitude.
For chiral quark models or in the Skyrmion picture of the nucleon,
the main contribution to the $E2$ strength stems from tensor correlations
between the pion cloud and the quark bag or meson exchange currents 
between the quarks. To observe a static deformation ($d$--state admixture) a
target with a spin of at least $3/2$ (e.g. $\Delta$ matter) would be
required. The only realistic alternative is to measure the transition $E2$ 
moment in the $\gamma N \rightarrow \Delta$ transition at resonance, 
or equivalently the
$E_{1+}^{3/2}$ partial wave amplitude in the $\Delta \rightarrow N \pi$
decay. The amplitudes in the $N \pi$ final state are usually denoted by
$E_{l\pm}^I$ and $M_{l\pm}^I$, where $E$ and $M$ are the electric and
magnetic multipoles, $I$ is the isospin and $l$ is the orbital angular 
momentum of the $N \pi$ system and the $\pm$ sign refers 
to its total angular momentum $J=l \pm 1/2$.

The experimental quantity of interest to compare with the different nucleon
models is the ratio $R_{EM} = E2/M1 = E_{1+}^{3/2}/M_{1+}^{3/2}$
of the electric quadrupole $E2$ to the magnetic dipole $M1$ amplitude in the 
region of the $\Delta(1232)$ resonance. In quark models with $SU(6)$ symmetry, 
for example the MIT bag model, $R_{EM} = 0$ is predicted.
Depending on the size of the hyperfine interaction and the bag radius, broken
$SU(6)$ symmetry leads to $-2\% < R_{EM} < 0$~\cite{Kon80,Ger81,Dre84,Cap92}. 
Larger negative values in the range $-6\% < R_{EM} < -2.5\%$ have been 
predicted by Skyrme models~\cite{Wir87} while results from chiral bag 
models~\cite{Ber88} give values in the range  
$-2\%$ to $-3\%$. The first Lattice QCD result is $R_{EM} = (+3 \pm
9)\%$~\cite{Lei92} and a quark model with exchange currents yields values 
of about $- 3.5 \%$~\cite{Buc97}.

The determination of the quadrupole strength $E2$ in the region of the
$\Delta(1232)$ resonance has been the aim of a considerable
number of experiments and theoretical activities in the last few years.
Very recently, new experimental results have been published for the
differential cross section and photon asymmetry of pion photoproduction
off the proton from the Mainz Microtron MAMI and the laser
backscattering facility LEGS at Brookhaven National
Laboratory, with the results $R_{EM} = - (2.5\pm0.2_{stat}\pm0.2_{sys})\%$
from the Mainz group~\cite{Bec97} and
$R_{EM} = - (3.0\pm0.3_{stat + sys}\pm0.2_{mod})\%$ from the LEGS
group~\cite{Bla97}. These new $R_{EM}$ results have started intense
discussions about the correct way to extract the $E2/M1$ ratio from the new
experimental data. In particular the large variation in the $R_{EM}$ values
obtained in theoretical analyses of these data at 
RPI~\cite{Dav97} ($R_{EM} = - (3.2\pm0.25)\%$), 
VPI~\cite{Wor97} ($R_{EM} = - (1.5\pm0.5)\%$) 
and Mainz~\cite{Han96} ($R_{EM} = - (2.5\pm0.1)\%$) was quite unsatisfactory.

In this paper we present the MAMI $p(\vec{\gamma},p)\pi^0$ and
$p(\vec{\gamma},\pi^+)n$ differential cross sections and photon asymmetries
and discuss in detail different analyses to extract the $s$-- and $p$--
partial wave amplitudes and the $E2/M1$ ratio.
In Section II we briefly describe the experimental setup and show a selection 
of the measured cross sections and photon asymmetries  for both pion
production channels from the proton in Section III. The essential ingredients 
of a multipole analysis 
are outlined in Section IV. Three different analyses of our data to extract 
the $s$-- and $p$--partial wave amplitudes and the $E2/M1$ ratio are 
discussed in Sections V and VI. We conclude with a summary and outlook 
in Section VII.

\section {\bf Experimental Setup and Data Analysis}

Since the experimental setup used for this measurement 
was described in detail in Ref. \cite{Cra96}, we will restrict the present 
discussion to the main features of the experiment.
Linearly polarized photons were produced by coherent bremsstrahlung
in a $100 ~\mu m$ thick diamond crystal~\cite{Loh94,Sch95}.
The photon energy was determined by the Glasgow tagging spectrometer at
the Mainz Microtron MAMI~\cite{Her90},
which in a 352 channel focal plane detector analyses the momentum of the
electron that has radiated the bremsstrahlung photon~\cite{Ant91}.
This detector system is able to energy tag photons in the range from
$50$ to $800 {\rm MeV}$  with a resolution of about $2{\rm MeV}$ \cite{Hal95}. 
The
collimation of the photon beam yielded a tagging efficiency of
about $55 \%$ for incoherent bremsstrahlung. To continuously monitor the 
tagging efficiency and the photon 
polarization, a pair detector was used downstream of the hadron detector
DAPHNE. This pair detector consists of a $0.5$ mm thick Cu--converter
followed by two $2$ mm thick plastic scintillators operated in coincidence.
Its efficiency ($\epsilon \simeq 3 \%$ for photons) was regularly checked 
with a
lead glass detector ($\epsilon \simeq 100 \%$ for photons) in calibration 
runs at
low beam intensity. The photon polarization was determined from the photon
spectrum measured by the tagging spectrometer in coincidence with the pair
detector and with the aid of theoretical calculations~\cite{Ram98}. 
The quality 
of these calculations was tested by an absolute measurement of the photon
polarization using coherent $\pi^0$ photoproduction on $^4He$ as a
polarimeter reaction with an analyzing power $A = 100 \%$. Excellent 
agreement
between calculations and experiment was found~\cite{Kra98} and in this way, 
both the photon polarization and the photon flux could be determined with 
an absolute precision of better than $\pm 2 \%$.

The liquid hydrogen cryogenic target was contained in a $43$ mm diameter,
$275$ mm long Mylar cylinder with a wall thickness of $0.1$ mm. The target
density was stabilized and determined to an accuracy of $\pm 0.5\%$
by means of an an automatic pressure and temperature control system.

The reaction products, the recoil proton from
$\gamma p \rightarrow p \pi^0$ and the pion from $\gamma p \rightarrow n
\pi^+$, were detected using the large acceptance detector DAPHNE 
$(21^0 \le \theta \le 159^0,~0^0 \le \phi \le 360^0)$ built by the
CEA/DAPNIA--SPhN at Saclay and the INFN-sezione di Pavia~\cite{Aud91}.
Good definition of the charged particle tracks was obtained from the
central vertex detector consisting of $3$ coaxial 
cylindrical multiwire proportional chambers providing a polar angular 
resolution of $\Delta\theta \le 1^o$ 
FWHM and an azimuthal resolution $\Delta\phi \le 2^o$ FWHM. 
This vertex detector is
surrounded by a segmented $\Delta E-E- \Delta E$ plastic scintillator 
telescope with successive thicknesses of $10$ mm, $100$ mm and $5$ mm 
respectively. 
The outermost layer is a lead--aluminium scintillator sandwich designed to
enhance the $\pi^0$ detection efficiency and to provide additional energy 
loss information for charged particles.

In the first step of the analysis those events were selected, that had only
one charged trajectory with its polar angle in the range $21^0 \le \theta
\le 159^0$. After this cut, the basic task of the data analysis was to
identify the pions and protons. This separation was performed by using a 
range method as described in detail in Ref. \cite{Bra94}, which 
simultaneously uses all
of the measured energy losses in the scintillator layers of DAPHNE to
identify charged hadrons ($\pi^+$, $p$) and to determine their energy. A
restriction made on the vertex position, defined as the point on the
reconstructed track, which lies closest to the detector axis, ensured a
complete rejection of particles coming from the target windows and walls. As
a result no empty target subtraction was needed. 

The main contributions to the systematic error in the determination of the
unpolarized differential cross section are due to uncertainties in the photon
flux ($\pm 2 \%$), the target density ($\pm 0.5 \%$) and ($\pm 2 \%$) for
the proton/pion separation with the range method.

\section {\bf Experimental Results}

In single pion photoproduction, the differential cross section for linearly 
polarized photons and unpolarized targets is given by the expression
\begin{equation}
\frac{d\sigma(\theta,\phi)}{d\Omega} = \frac{d\sigma_0(\theta)}{d\Omega}
(1 ~-~ \Sigma(\theta) \,cos(2\phi)) \, ,
\end{equation}
where $\Sigma$ is the photon asymmetry and $\theta$ and $\phi$ are the 
polar and azimuthal angles of the pion with respect to the beam direction.
Since DAPHNE has full 2$\pi$ azimuthal coverage it allows a direct 
measurement of the $\phi$ dependence of the differential cross 
section, and therefore, the determination of $\Sigma$ and the unpolarized 
cross section $d\sigma_{0}/d\Omega$ at the same time.

\begin{figure}
\centerline{\psfig{figure=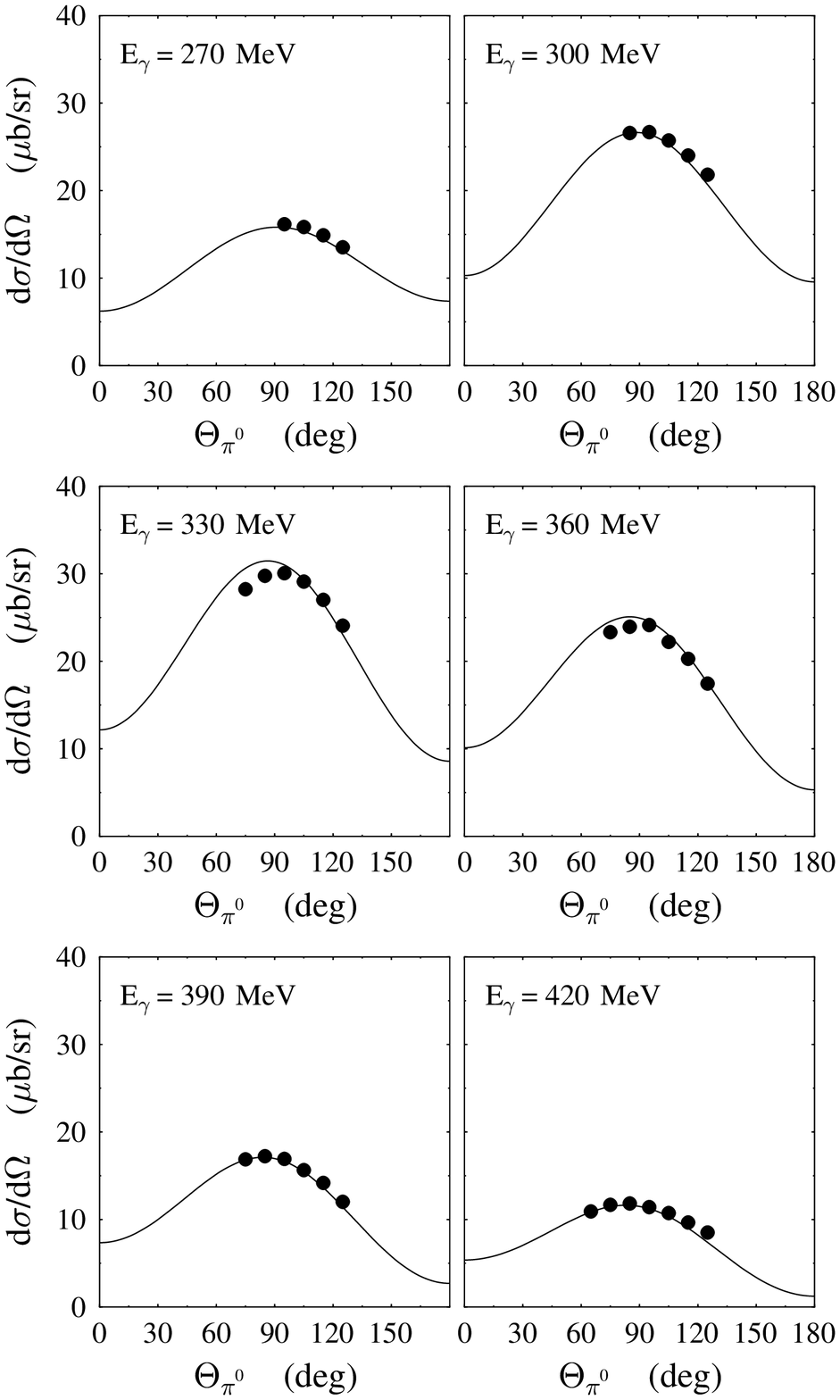,height=20cm}}
\caption{The differential cross section $d\sigma/d\Omega$
         for $p(\gamma,p)\pi^0$
         at $6$ different photon energies. The solid line shows the 
         result of the energy dependent multipole analysis.}
\label{fig:pi0d}
\end{figure}

\begin{figure}
\centerline{\psfig{figure=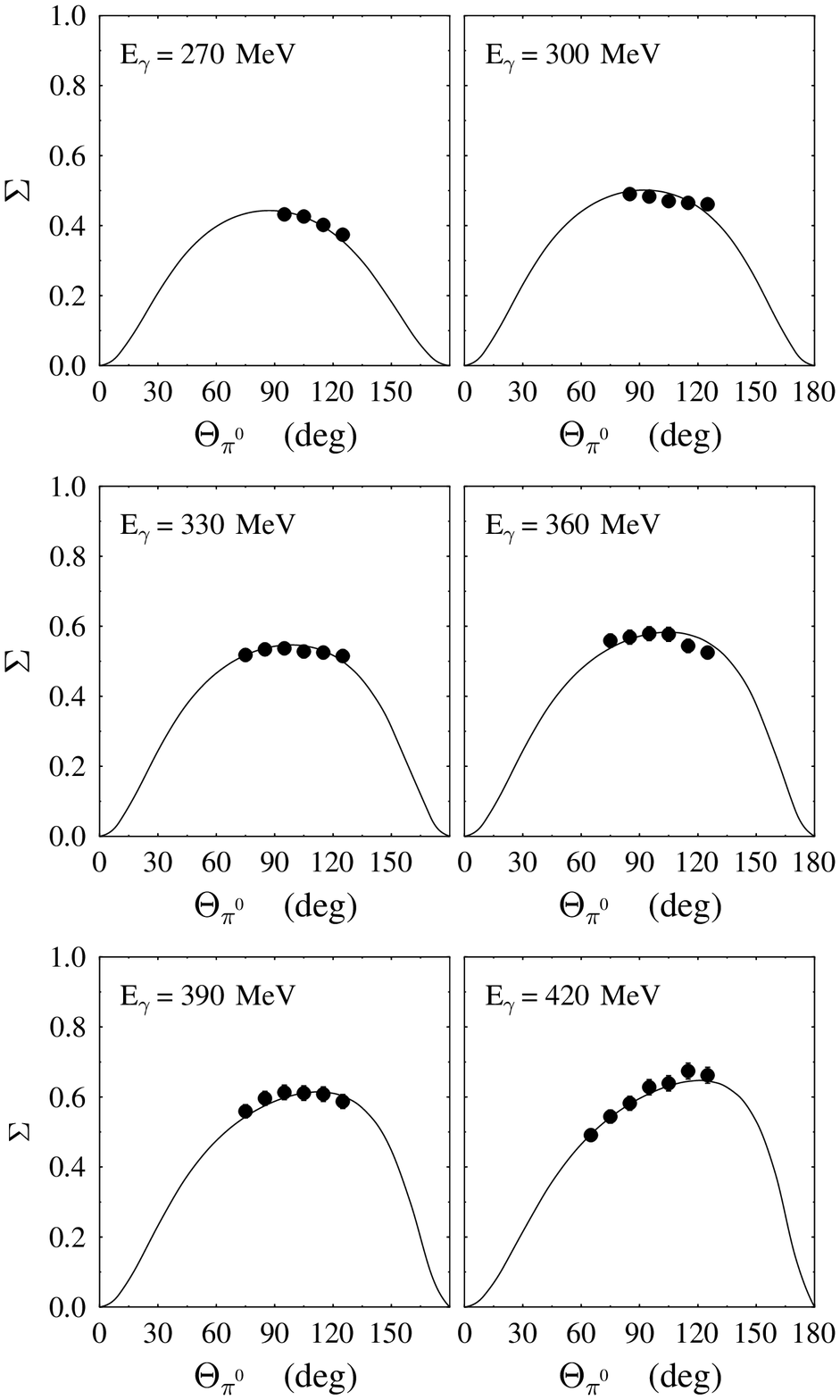,height=20cm}}
\caption{The photon asymmetry $\Sigma$ for $p(\vec{\gamma},p)\pi^0$
         at $6$ different photon energies. The solid line shows the 
         result of the energy dependent multipole analysis.}
\label{fig:pi0s}
\end{figure}

\begin{figure}
\centerline{\psfig{figure=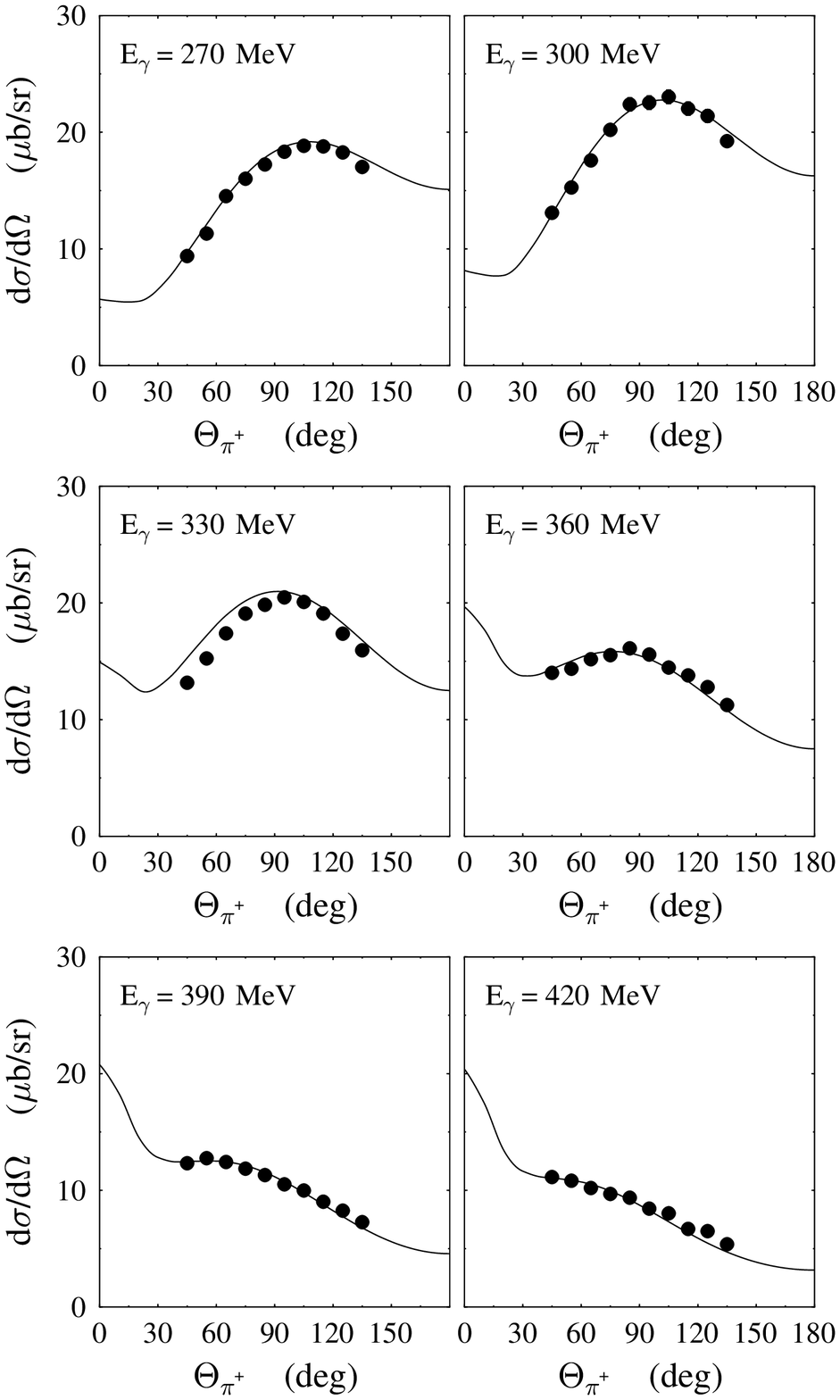,height=20cm}}
\caption{The differential cross section $d\sigma/d\Omega$
         for $p(\gamma,\pi^+)n$ at $6$ different photon energies. 
         The solid line shows the result of the energy dependent multipole 
         analysis.}
\label{fig:pipd}
\end{figure}

\begin{figure}
\centerline{\psfig{figure=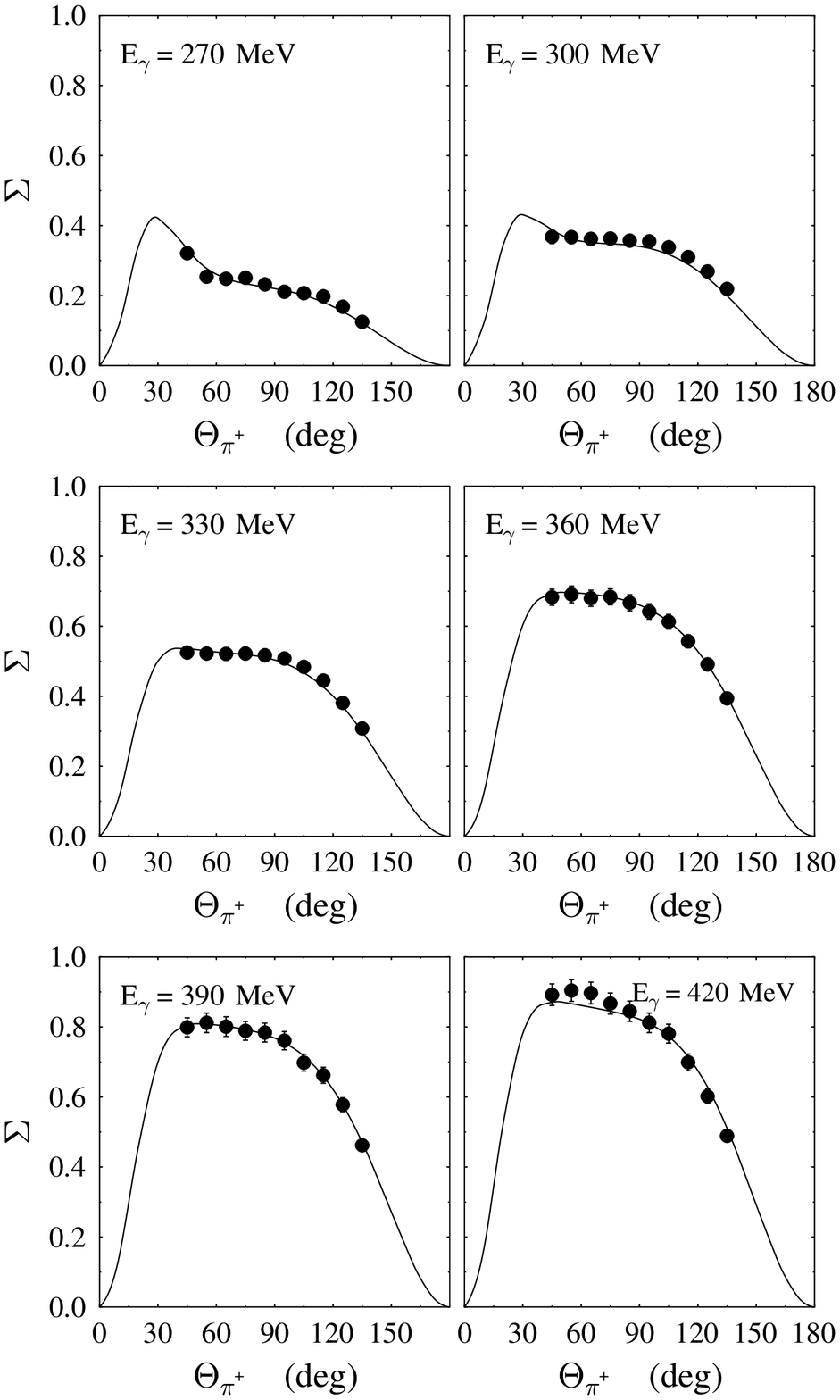,height=20cm}}
\caption{The photon asymmetry $\Sigma$ for $p(\vec{\gamma},\pi^+)n$
         at $6$ different photon energies. The solid line shows the 
         result of the energy dependent multipole analysis.}
\label{fig:pips}
\end{figure}

A selection of the measured cross sections and photon asymmetries in the 
cm-frame for the 
$\gamma p \rightarrow p \pi^0$ and $\gamma p \rightarrow n \pi^+$ reaction 
channels 
\cite{Bec97,Kra96} are shown in Figs.~\ref{fig:pi0d}--\ref{fig:pips}.
This is the first data set for which
the two observables $d\sigma_{0}/d\Omega$ and $\Sigma$ have been measured
simultaneously at all angles and photon energies for both pion production
channels from the proton. The measurement covers the whole $\Delta(1232)$
resonance region in $16$ energy bins between $E_{\gamma}=270 {\rm MeV}$ and
$420 {\rm MeV}$, where $E_{\gamma}$ is the tagged photon energy in the 
lab-frame. The angular distributions for $d\sigma_{0}/d\Omega$ and
$\Sigma$ were measured from $45^0$ up to $135^0$ for the
$n \pi^+$ channel.
In the $p \pi^0$ case the angular range varied from
$95^0$ to $125^0$ at $E_{\gamma}=270 {\rm MeV}$ to $65^0$ to $125^0$ for the
highest energy point at $E_{\gamma}=420 {\rm MeV}$.
In Figs.~\ref{fig:pi0d}--\ref{fig:pips} the result of a fixed--t
dispersion relation analysis~\cite{Han98}, which will be discussed below,
is shown as a solid line. Our differential cross sections and asymmetries
are in good agreement with the dispersion relation analysis except for the
differential cross sections around $330 {\rm MeV}$ where we find a slight
discrepancy.

\begin{figure}
\centerline{\psfig{figure=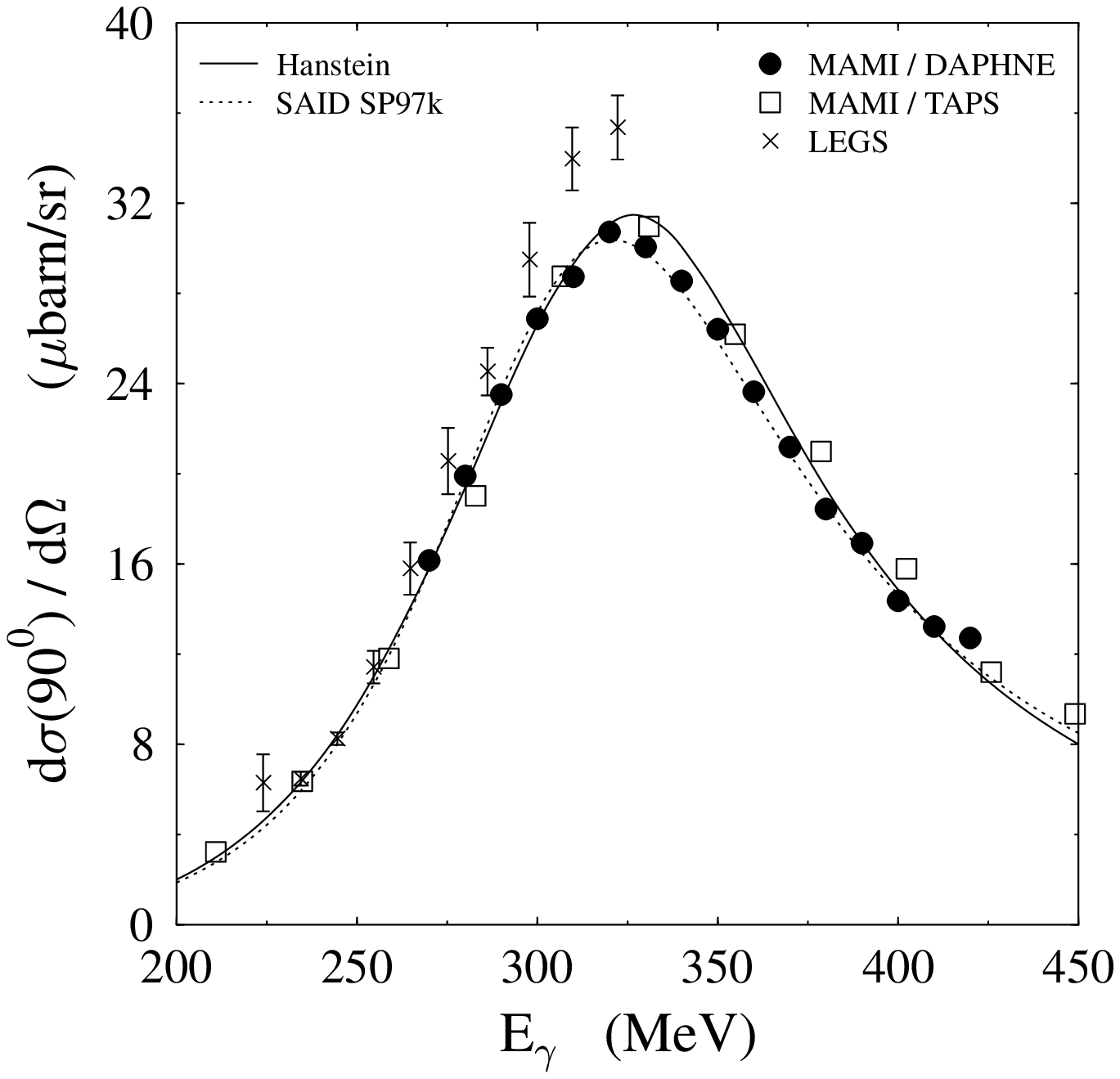,height=12.0cm}}
\caption{The energy dependence of the differential cross section
         at $\theta_{\pi^0} = 90^0$ for $p(\gamma,p)\pi^0$ 
         in comparison with our energy dependent multipole analysis (full
         line) and the VPI solution SP97k (dotted line). For the LEGS points
         the statistical error and part of the systematic error are combined
         to the net uncertainty bars \protect\cite{San98}. The error bars on
         the MAMI/DAPHNE points \protect\cite{Kra96} and the MAMI/TAPS results
				 \protect\cite{Hae96} are only statistical. The systematic scale
         uncertainties are $\sim 2 \%$ for LEGS, $\sim 4 \%$ for MAMI/DAPHNE 
				 and $\sim 6 \%$ for MAMI/TAPS.}
\label{fig:pa90}
\end{figure}

\begin{figure}
\centerline{\psfig{figure=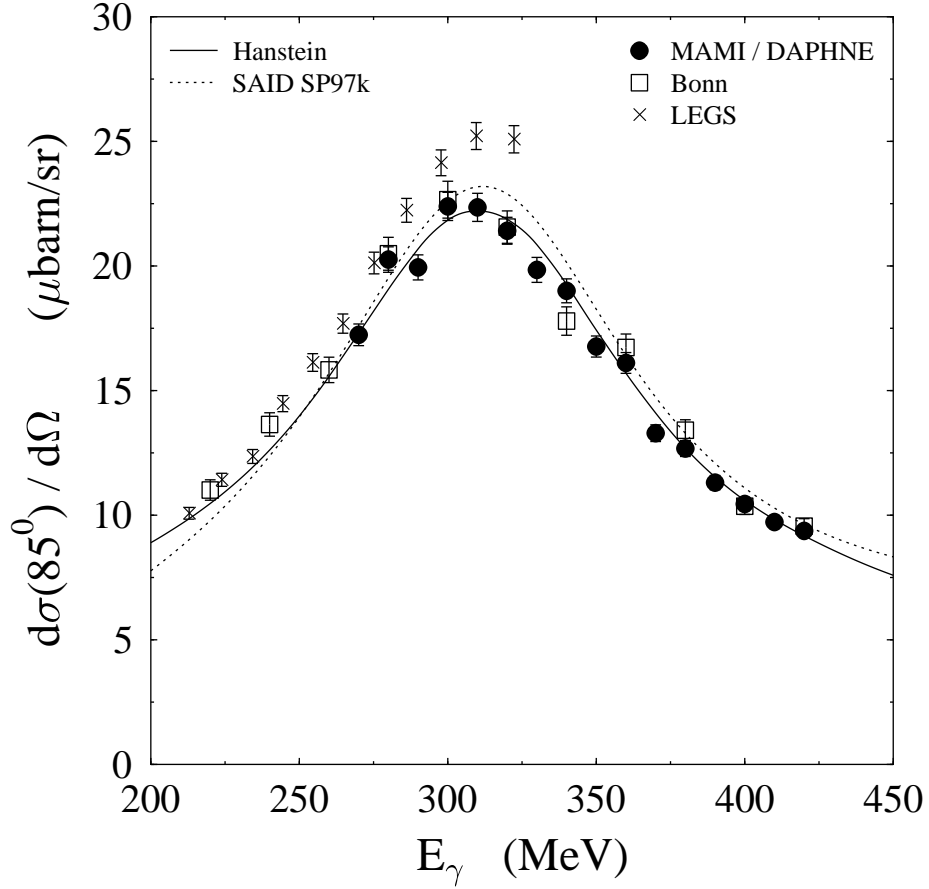,height=12.0cm}}
\caption{The energy dependence of the differential cross section
         at $\theta_{\pi^+} = 85^0$ for $p(\gamma,\pi^+)n$
         in comparison with our energy dependent multipole analysis (full
         line) and the VPI solution SP97k (dotted line). For the LEGS points
         the statistical error and part of the systematic error are combined
         to the net uncertainty bars \protect\cite{San98}. The error bars on
         the MAMI points \protect\cite{Kra96} and the Bonn results
				 \protect\cite{Fis72} are only statistical. The systematic scale
         uncertainties are $\sim 2 \%$ for LEGS, $\sim 4 \%$ for MAMI and 
				 $\sim 6 \%$ for Bonn.}
\label{fig:pa85}
\end{figure}

Figs.~\ref{fig:pa90} and~\ref{fig:pa85} compare the
energy dependence of the present
MAMI results and the recent LEGS results \cite{Bla97}
for $\gamma p \rightarrow p \pi^0$ at $\theta_{\pi^0} = 90^0$
and $\gamma p \rightarrow n \pi^+$ at $\theta_{\pi^+}=85^0$.
For the LEGS results the statistical error and the angle and energy
dependent error have been evaluated point by point and combined to the net
uncertainty bars, which are shown in Figs. ~\ref{fig:pa90}
and~\ref{fig:pa85}. The error bars on the MAMI points are only statistical.
The systematic scale uncertainties are $\sim 3 \%$ for the MAMI results and
$\sim 1 \%$ for the LEGS results.  
There is an energy dependent discrepancy of up to $15 \%$
between the absolute differential cross sections. 
For both $\gamma p \rightarrow p \pi^0$
and $\gamma p \rightarrow n \pi^+$, the LEGS cross
sections start to rise above the MAMI data at $E_{\gamma}=280 {\rm MeV}$,
reaching $\simeq 15 \%$ higher values at the highest energy LEGS data 
point at $E_{\gamma}=323 {\rm MeV}$.
Since the difference is energy dependent
it will affect the absolute values of the resonant partial waves
$M_{1+}^{3/2}$ and $E_{1+}^{3/2}$ (see discussion below).

\begin{figure}
\centerline{\psfig{figure=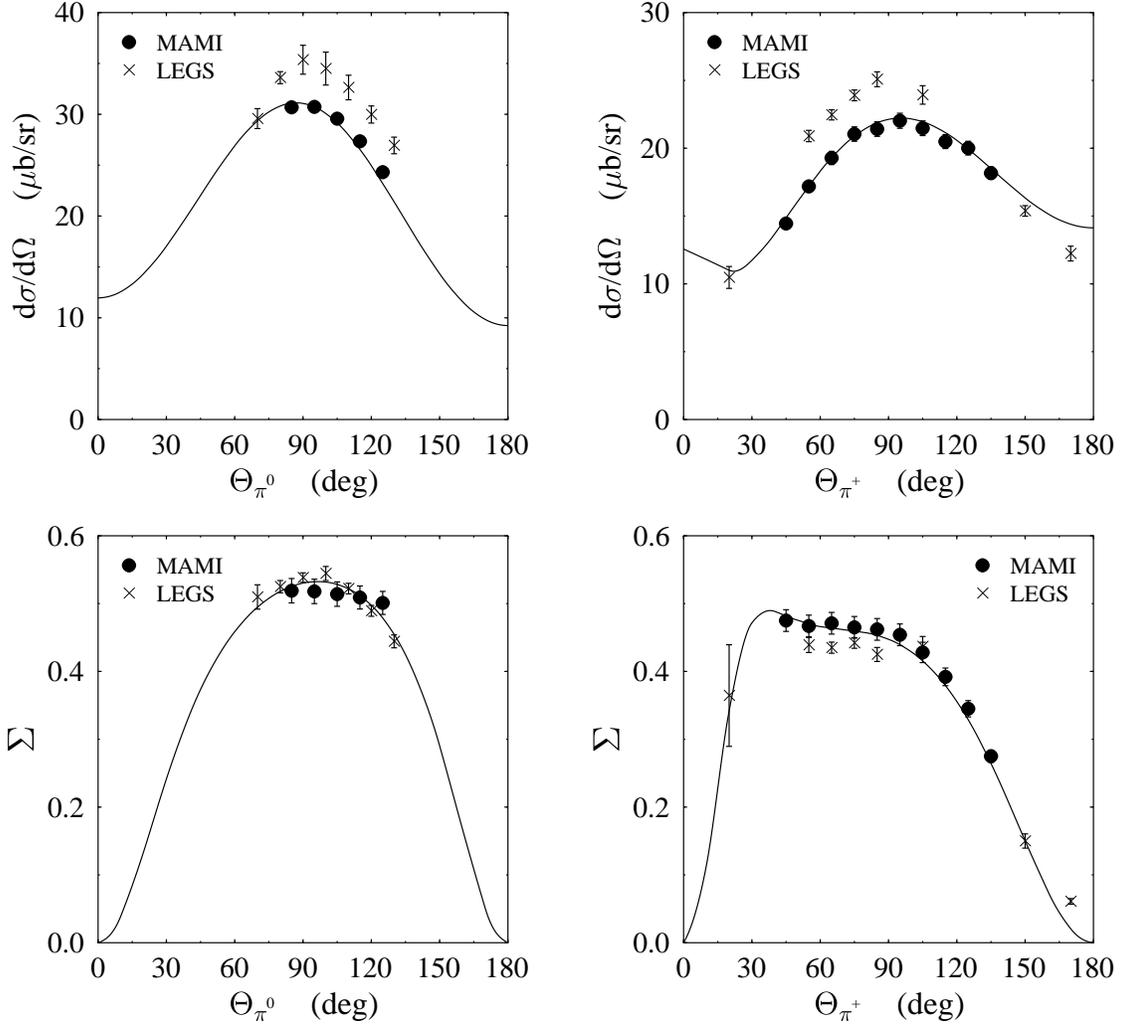,width=15cm}}
\caption{The differential cross section $d\sigma/d\Omega$ and
         the photon asymmetry $\Sigma$ for $p(\vec{\gamma},p)\pi^0$
         (left) and $p(\vec{\gamma},\pi^+)n$ (right) at
	       $E_{\gamma} = 320 {\rm MeV}$. The MAMI data are compared to
				 the results of LEGS and our energy dependent multipole analysis
         (full line). For the LEGS points
         the statistical error and part of the systematic error are combined
         to the net uncertainty bars \protect\cite{San98}. The error bars on
         the MAMI points \protect\cite{Kra96} are only statistical. 
				 The systematic scale uncertainties for $d\sigma/d\Omega$ 
				 are $\sim 2 \%$ for LEGS and $\sim 4 \%$ for MAMI and $\sim 2 \%$ 
				 for the photon asymmetry of MAMI.}
\label{fig:wq}
\end{figure}

In Fig.~\ref{fig:wq} the LEGS and MAMI angular
distributions for the differential cross
section and the photon asymmetry are compared at $E_{\gamma}=320 {\rm MeV}$
for  $p(\vec{\gamma},p)\pi^0$ and $p(\vec{\gamma},\pi^+)n$.
One sees that the differential cross sections differ not only in 
absolute magnitude, but also in angular shape.
The two most forward angular points at $\theta_{\pi^0}=70^0$ and $80^0$
of the LEGS $p\pi^0$ differential cross section drop faster than 
the MAMI results. The same behavior is seen for the $n\pi^+$ cross 
section at $\theta_{\pi^+}=20^0$, $150^0$ and $170^0$.
This difference in the shape of the differential cross sections for the 
two data sets will play an important role in the discussion
of the non--Born contribution for partial waves with $l_{\pi} \ge 2$.
On the other hand, the photon asymmetry agrees well for the $p\pi^0$
channel and shows only small differences for the $n\pi^+$ data,
where the MAMI results are slightly above the LEGS data.

\section {\bf Multipole Analysis of the MAMI data}

\subsection {\bf General Aspects}

In $\gamma N \rightarrow N \pi$ both incident particles and the final 
nucleon have two spin states yielding in eight degrees of freedom. 
From parity conservation this is reduced to a total number of
four complex amplitudes to describe the reaction.
Allowing for one arbitrary phase factor, there are therefore
seven independent physical quantities which need to be measured at 
any setting of $E_{\gamma}$ and $\theta_{\pi}$.
Different sets of amplitudes are used in the literature, e.g. 
the helicity
amplitudes ($H_1$, $H_2$, $H_3$, $H_4$) introduced by Jacob and Wick
\cite{Wic59} or the Pauli amplitudes ($F_1$, $F_2$, $F_3$, $F_4$)
by Chew, Goldberger, Low and Nambu \cite{CGLN57}.
While the observables of single pion photoproduction are more elegantly 
expressed by the helicity amplitudes, the Pauli
amplitudes are particularly suited for a decomposition into partial 
waves~\cite{Ras89,Dre92}.

A complete database for pion photoproduction (the ``complete'' experiment) 
requires at least $8$ independent
observables to specify the multipole amplitudes to all orders in $l_{\pi}$
\cite{Bar75,Chi97,Wor96}. Such complete information is not available at
present and standard multipole analyses have to rely on the
differential cross section $d\sigma/d\Omega$ and the
three single polarization observables $\Sigma$
(photon asymmetry), $P$ (recoil nucleon polarization) and $T$ (target
asymmetry).
For pion photoproduction from threshold up to the
$\Delta(1232)$ resonance region, these four observables provide
sufficient conditions for a complete database, if the higher partial
waves $l_{\pi}\ge2$ can be adequately represented by the Born contributions.
This approach is expected to be appropriate up to the region of the higher 
resonances
($E_{\gamma} \simeq 600 {\rm MeV}$) and even there may affect only certain 
multipoles (e.g. $D_{13}(1520) \leftrightarrow E_{2^-}, ~M_{2^-}$).
Such arguments were used by V. Grushin \cite{Gru75} to analyse the Kharkov
data~\cite{Kha80}
($d\sigma/d\Omega$, $\Sigma$, $P$ and $T$) for $\gamma p \rightarrow p \pi^0$
and $\gamma p \rightarrow n \pi^+$. In their analysis both the real and 
imaginary parts of the $s$ and $p$ wave amplitudes could be determined 
for the first time independently of the
pion-nucleon phase shifts $\delta_{IJ}^{\pi N}$. 
The mean difference between the $\delta_{33}^{\gamma, \pi N}$ phase from the 
Kharkov analysis and the known pion--nucleon scattering phase 
$\delta_{33}^{\pi N}$ was only $- (2.3 \pm 0.5)^0$ 
over the entire energy range $E_{\gamma} = 250 - 500 {\rm MeV}$. 

The $p\pi^0$ and $n\pi^+$ multipole amplitudes ${\cal M}_{l\pm}$ 
(${\cal M}_{l\pm}$ stands for $E_{0^+}$, $M_{1^-},\ldots $) are related 
to the isospin $1/2$ and $3/2$ components 
${\cal M}_{l\pm}^{1/2}$ and ${\cal M}_{l\pm}^{3/2}$ by
\begin{eqnarray}
{\cal M}_{l\pm}(p\pi^0) &=& {\cal M}_{l\pm}^{1/2} + \frac{2}{3}{\cal M}_{l\pm}^{3/2} \, \\ 
{\cal M}_{l\pm}(n\pi^+) &=& \sqrt{2}({\cal M}_{l\pm}^{1/2} - \frac{1}{3}{\cal M}_{l\pm}^{3/2}) \, .
\end{eqnarray}
The multipole amplitudes ${\cal M}_{l\pm}^I$ are complex functions of the
c.m. energy W. Below the two pion production threshold 
($E_{\gamma} \simeq 310 {\rm MeV}$), the
Fermi--Watson theorem~\cite{Wat54} allows one to express the phases of the 
complex multipole amplitudes by the corresponding pion--nucleon scattering 
phase shifts $\delta_{l\pm}^I$
\begin{equation}
{\cal M}_{l\pm}^{I} = |{\cal M}_{l\pm}^{I}| exp{(i\delta_{l\pm}^I + n\pi)} \, ,
\end{equation}
where $I$ can be $1/2$ and $3/2$ and $n$ is an integer. However, even 
at energies of about $400 {\rm MeV}$, the $\pi N$ inelasticities in 
the $P_{33}$ partial wave are very small, which suggests that the 
Fermi--Watson theorem can be applied well above the two pion threshold. 

There exist two basic methods to extract the multipole amplitudes 
from the database,
the ``energy independent'' and the ``energy dependent'' approach.
In the energy independent approach, each energy is investigated
independently by use of  standard $\chi^2$ minimization techniques, the
fit parameters being the real and imaginary parts of
the multipole amplitudes ${\cal M}_{l\pm}^{I}$.
Below the two pion production threshold the Fermi--Watson theorem is used, 
which reduces the number
of the necessary observables for a complete experiment by a factor of two, 
because only the absolute values $|{\cal M}_{l\pm}^{I}|$ of the partial 
wave amplitudes need to be determined from the fit.
In the energy dependent approach, the data at all energies are
analyzed simultaneously. Either an energy dependent parametrization of 
the partial wave amplitudes must be assumed, or as in our case, the energy
dependence is taken from dispersion relations.
The principal advantage of this method is that continuity is built in 
from the beginning and systematic errors tend to cancel out. In some cases
one or the other of these two approaches is to be preferred. If the data are
closely spaced in energy and cover both differential cross sections and
polarization observables (``complete experiment'') at each energy, the
energy independent approach is more advantageous.
If on the other hand the data are widely spaced and only few polarization
observables are available, the energy dependent approach is the better
one. This approach is also useful if the general resonance
structure is already known and the main interest is to obtain the 
small partial wave amplitudes. However, the experimental data has to cover 
the complete energy range of the dominating resonances in order to allow a 
reliable separation. 
For instance in the case of the $M_{1+}$ amplitude in the 
$\gamma N \rightarrow \Delta(1232)$ transition, one has to cover the complete
resonance region ($250 - 450 {\rm MeV}$) to get the $M_{1+}$ multipole as 
precisely as possible \cite{Don73,Gru75}.

\subsection {\bf Higher partial waves}

A general problem common to both the energy dependent and the energy
independent approach is to decide at which
value of the angular momentum $l_{\pi}$ the partial wave expansion should be
truncated in the fit and how to approximate the higher partial waves.
The maximum value $n_{max}$ that the data can determine is found by fitting 
the angular distributions to a Legendre series or, equivalently, a power 
series expansion in $cos\theta$

\begin{eqnarray}
\frac{d\sigma}{d\Omega} & = & \frac{q}{k} \sum_{n=0}^{n_{max}} A^n cos^n\theta\, , \\ 
\Sigma \frac{d\sigma}{d\Omega} \frac{1}{sin^2\theta} & = & \frac{q}{k} 
\sum_{n=0}^{n_{max}} A_{\Sigma}^n cos^n\theta\, , \\
T \frac{d\sigma}{d\Omega} \frac{1}{sin\theta} & = & \frac{q}{k} 
\sum_{n=0}^{n_{max}} A_T^n cos^n\theta\, , \\
P \frac{d\sigma}{d\Omega} \frac{1}{sin\theta} & = & \frac{q}{k} 
\sum_{n=0}^{n_{max}} A_P^n cos^n\theta\, .
\end{eqnarray}
The experimentally accessible polynomial coefficients $A^n$ 
are quadratic or bilinear products of the electric $E_{l\pm}$ and magnetic 
$M_{l\pm}$ multipole amplitudes. 

In this section we examine the sensitivity to higher partial wave
contributions of the coefficients $A^n$ and $A_{\Sigma}^n$ extracted from
the present data. The difference of the 
cross sections for the photon polarization perpendicular and parallel to the
reaction plane, $d\sigma_{\bot}/d\Omega - d\sigma_{||}/d\Omega$,
is particularly sensitive to higher partial waves. With partial waves up 
to $l_{\pi}=2$ the difference
\begin{equation}
\frac{1}{2 sin^2\theta} \left(\frac{d\sigma_{\bot}}{d\Omega} - 
\frac{d\sigma_{||}}{d\Omega}\right)
= \Sigma \frac{d\sigma}{d\Omega} \frac{1}{sin^2\theta} = \frac{q}{k}
(A_{\Sigma} + B_{\Sigma} cos\theta + C_{\Sigma} cos^2\theta)
\end{equation}
has three polynomial coefficients $A_{\Sigma}$, $B_{\Sigma}$  and 
$C_{\Sigma}$ with
\begin{eqnarray}
A_{\Sigma} &\simeq& A_{\Sigma}(s_{wave},p_{wave}) + {\rm Re} \left[E_{0+}d^*\!\!_{wave}  \right] +  
| d_{wave} |^2  \; , \\
B_{\Sigma} &\simeq& {\rm Re} \left[ (M_{1+}-M_{1-}) d^*_{wave}  \right] \; , \\
C_{\Sigma} &=& | d_{wave} |^2 \; . 
\end{eqnarray}
In the case that only the $s$ wave ($E_{0+}$) and $p$ waves 
($M_{1+},~M_{1-}$ and  $E_{1+}$) contribute, 
this difference would be proportional to $A_{\Sigma}$ and therefore 
independent of the pion angle $\theta$. The multipole dependence of 
the coefficients $A_{\Sigma}$, $B_{\Sigma}$ and $C_{\Sigma}$ is described in
Appendix A.

Fig.~\ref{fig:sp} shows the difference of $d\sigma_{\bot}/d\Omega -
d\sigma_{||}/d\Omega$  
for both pion reaction channels on the proton
at our lowest photon energy $E_{\gamma} = 270 {\rm MeV}$, at
$E_{\gamma} = 350 {\rm MeV}$ and at our highest energy 
$E_{\gamma} = 420 {\rm MeV}$. In the $p \pi^0$ channel one recognizes 
only above the resonance a small 
deviation from the constant behavior. This is due entirely to the Born 
contribution to the $B_{\Sigma}$ coefficient, i.e. the 
interference between the real part of the dominant $M_{1+}$ amplitude and 
the real part of the $d$ wave amplitudes, for example in terms like 
${\rm Re}(M_{1+} - M_{1-}) {\rm Re}E_{2-}$. 
Such contributions become extremely small at resonance since 
${\rm Re}(M_{1+} - M_{1-}) \simeq 0$.

The behavior of the difference, $d\sigma_{\bot}/d\Omega -
d\sigma_{||}/d\Omega$, for
the $\vec{\gamma} p \rightarrow p \pi^0$ data from the LEGS 
collaboration is different. Their observed angular dependence
seems to require a sizeable non--Born contribution from $d$ and
$f$ waves. The origin of the different angular dependence
of the LEGS results for $d\sigma_{\bot}/d\Omega - d\sigma_{||}/d\Omega$ 
arises from the different shape of their unpolarized cross section data.
The LEGS photon asymmetries alone are well described by our multipole
fit which takes Born contributions in higher partial waves of 
$l_{\pi} \ge 2$ into account. 

In contrast the difference of 
$d\sigma_{\bot}/d\Omega - d\sigma_{||}/d\Omega$
for $\vec{\gamma} p \rightarrow n \pi^+$ shows a strong angular dependence 
at all energies. The main reason is that in charged pion photoproduction
the pion pole Born graph leads to significant contributions of higher
partial waves.

\begin{figure}
    \centerline{\psfig{figure=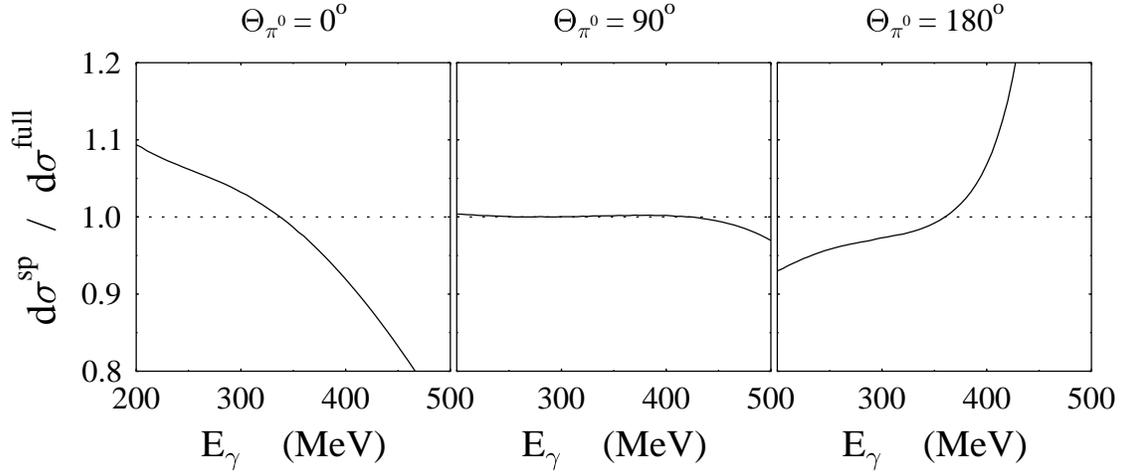,width=15cm}}
    \caption{The contribution of higher partial waves ($l_{\pi} \ge 2$) to the
             differential cross section for $\gamma p \rightarrow p \pi^0$ at
             $\theta_{\pi}=0^0$, $\theta_{\pi}=90^0$ and $\theta_{\pi}=180^0$.
	           The solid line is the ratio ($d\sigma^{sp} / d\sigma^{full} $) 
             of the differential cross section for only $s$ and $p$ waves 
             to the differential cross section including higher partial wave 
	           contributions (Born contribution for $l_{\pi} \ge 4$). 
             The curves are obtained from SAID \protect\cite{Sai96} 
             solution SM95.}
\label{fig:saidsp}
\end{figure}

The sensitivity of the differential cross section to higher partial waves is
most pronounced at the extreme forward and backward angles. 
Taking partial waves up to $l_{\pi}=2$, the differential cross section is
\begin{eqnarray}
\frac{d\sigma}{d\Omega} = \frac{q}{k}( A + B cos(\theta) + C cos^2(\theta) +
D cos^3(\theta) + E cos^4(\theta))  
\end{eqnarray}
with five coefficients of the form
\begin{eqnarray}
A &\simeq& A(s_{wave},p_{wave}) +  {\rm Re} \left[E_{0+}d^*\!\!_{wave}  \right] +  
| d_{wave} |^2  \; , \\
B &\simeq& B(s_{wave},p_{wave}) + {\rm Re} \left[ (M_{1+}-M_{1-}) d^*_{wave}  \right] \; , \\
C &\simeq& C(s_{wave},p_{wave}) + {\rm Re}\left[ E_{0+} d^*\!\!_{wave} \right] + |
d_{wave} |^2 \; , \\
D &\simeq& {\rm Re} \left[ (M_{1+}-M_{1-}) d^*\!\!_{wave} \right] \; , \\
E &=& | d_{wave} |^2 \; . 
\end{eqnarray}
The effect of the $d$ waves is largest for the coefficients B and D, because
of an interference term between the large $M_{1+}$ amplitude and the 
$d$ waves. However, 
the contributions of these terms can be neglected at the top of the resonance 
($\delta_{33} = 90^0$). As an example this is illustrated for the combination
\begin{eqnarray}
{\rm Re} \left[ (M_{1+}-M_{1-}) E_{2-}^* \right] &=& 
{\rm Re} (M_{1+}-M_{1-}) {\rm Re} E_{2-} + 
{\rm Im} (M_{1+}-M_{1-}) {\rm Im} E_{2-} \; ,
\end{eqnarray}
where the first term vanishes, because ${\rm Re} (M_{1+}-M_{1-})$ passes
through zero at the resonance position ($\delta_{33} = 90^0$) and
the second term can be neglected, because ${\rm Im} E_{2-}$ is small due
to the small d-wave phase.
Fig.~\ref{fig:saidsp} illustrates the sensitivity of the differential cross
section to higher partial waves. The solid line shows the ratio
($d\sigma^{sp} / d\sigma^{full} $) of the differential cross section for
only $s$-- and $p$--wave contributions to the differential cross section
with higher partial waves. The calculated cross section ratio is shown
at the pion angles $\theta_{\pi^0} = 0^0,~90^0$ and $180^0$ in the energy
region between $200$ and $500 {\rm MeV}$. For this calculation~\cite{Sai96} the
SAID solution SM95 was used, in which partial waves up to $l_\pi=3$ 
were allowed to vary to fit the data set but the normal Born contribution
was assumed for $l_{\pi} \ge 4$. 
At $\theta_{\pi^0} = 90^0$, the contributions from the higher partial waves 
are far below $1\%$, because they arise only from an interference term 
with the small $s$ wave $E_{0^+}$, for example in the form  
${\rm Re} (E_{0^+}d^*_{wave})$ as in eq. 15.
Below and above the $\Delta(1232)$ resonance, however, contributions 
from $l_{\pi} \ge 2$ are of the order of $10 - 20 \%$ of the differential 
cross section at $0^0$ and $180^0$.
These contributions arise almost completely from the interference between the 
dominant $M_{1+}$ amplitude and the Born parts of the 
higher partial waves.
In our analysis there are no indications of significant non--Born 
contributions for higher partial waves ($l_{\pi} \ge 2$).  

\subsection {\bf Influence of systematic errors in the database}

An important issue for any multipole analysis is the question
of how to handle the systematic errors of the experimental
results for the different observables used in the database.
The ideal database should contain experimental results for both
$\gamma p \rightarrow p \pi^0$ and $\gamma p \rightarrow n \pi^+$, 
measured simultaneously with one setup
for all photon energies and covering the full range of polar angles 
to minimize energy and angular dependent systematic errors.
The analysis of such data would yield a consistent separation
of the isospin $1/2$ and $3/2$ parts from the partial waves of the
$p \pi^0$ and $n \pi^+$ reaction channels.
%
%\begin{eqnarray}
%M_{l\pm}(p\pi^0) = M_{l\pm}^{1/2} + \frac{2}{3}M_{l\pm}^{3/2} \\ 
%M_{l\pm}(n\pi^+) = \sqrt{2}(M_{l\pm}^{1/2} - \frac{1}{3}M_{l\pm}^{3/2}) \, .
%\end{eqnarray}
%
However, the isospin separation will fail if there is a considerable 
inconsistency, such as a normalization error, between the 
$\gamma p \rightarrow n \pi^+$ and
$\gamma p \rightarrow p \pi^0$ observables 
($d\sigma/d\Omega$, $\Sigma$, $T$, \ldots).
The existing pion photoproduction data, for example the SAID database 
are far from the requirements of the ideal case. Most of the 
experimental data are taken at one photon energy at a time, the angular
distributions are not measured simultaneously for all angles and, 
more importantly, the consistency between different measurements is less
good than their assigned errors would suggest.
When combining data from different experiments into one database, one has 
the difficult task to account properly for the angular  
and energy dependent systematic errors.
A common procedure is to combine the experimental systematic errors in 
quadrature with the statistical errors or to multiply all data from one set
with a systematic scale error ($\sigma_{sys}$) by a common factor $f$ 
while adding $(f-1)^2/\sigma_{sys}^2$ to the $\chi^2$~\cite{Ago94}.
The latter method allows for adjustments of an overall scale 
(angle independent systematic errors), but not for systematic effects 
in the shape of the angular distributions which 
is important for the determination of the small $E2$ amplitude.
As shown in the following paragraph, it is the shape of the 
differential cross section and the photon asymmetry near $\theta = 90^0$,
however, which is sensitive to the $E2$ amplitude.
Combining all existing $p\pi^0$ and $n\pi^+$ data in one database
regardless of their consistency will result
in mean values for the dominant partial waves ($M_{1+}$ and $E_{0+}$) but 
meaningless values for the small partial waves ($E_{1+}$ and $M_{1-}$).
Any reliable partial wave analysis for the small amplitudes
will require a careful selection of the experimental results used in the 
database. In practice one has to choose observables that are sensitive 
to the small amplitude from one experiment, 
or one has to use analyzing methods which 
are insensitive to energy and angle dependent systematic 
uncertainties.

\begin{figure}
\centerline{\psfig{figure=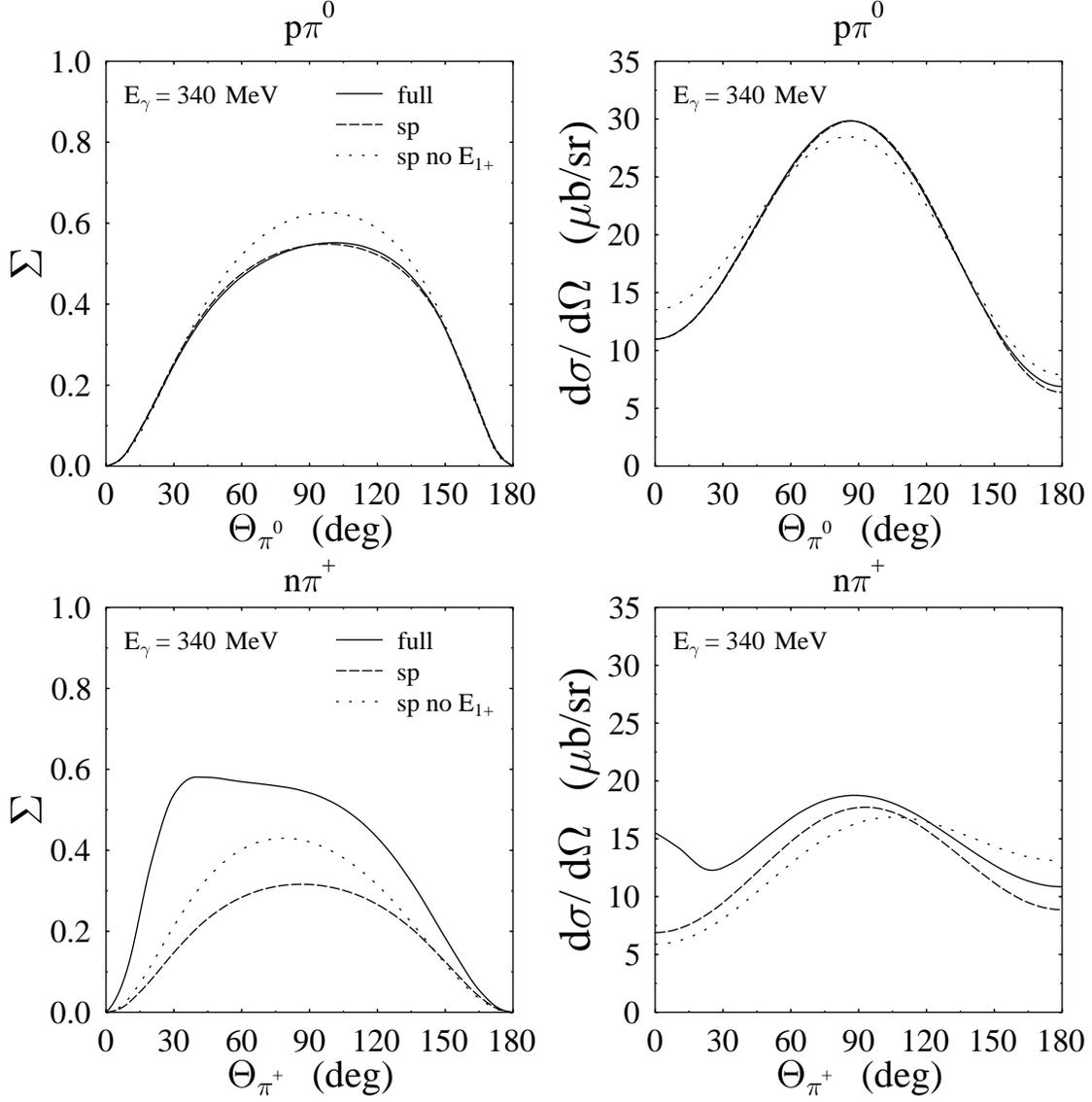,width=15.0cm}}
\caption{The sensitivity of differential cross sections 
         $d\sigma/d\Omega$ and photon asymmetries $\Sigma$ to the 
				 $E_{1+}$ multipole and higher partial waves ($l_{\pi} \ge 2$) 
				 for $p(\vec{\gamma},p)\pi^0$ and 
				 $p(\vec{\gamma},\pi^+p)n$ at $E_{\gamma}=340 {\rm MeV}$. The dashed 
				 lines are obtained with $s$ and $p$ waves only. The dotted 
				 lines excludes the $E_{1+}$ amplitude and the full lines show 
				 the results with all partial waves.}
\label{fig:sp340}
\end{figure}

Fig.~\ref{fig:sp340} shows $d\sigma/d\Omega$ and $\Sigma$ for different 
multipole contributions at the $\Delta(1232)$ resonance ($\delta_{33}=90^0$).
This illustrates the sensitivity to the higher partial waves ($l_{\pi} \ge 2$) 
and to the small $E_{1+}$ 
amplitude. The contribution of the higher partial waves to the 
$\gamma p \rightarrow n\pi^+$ channel is significant at all angles. 
For the differential cross section at $\theta_{\pi^+}=0^0$
and the photon asymmetry around $\theta_{\pi^+}=90^0$ the full calculation 
and the truncation to $s$ and $p$ waves differ by as much as a factor of two. 
The $\gamma p \rightarrow p \pi^0$ case is 
different and one cannot distinguish between the
full calculation and the result where only s and p waves are taken into
account. The main reason is that the pion pole term does not contribute 
in neutral pion photoproduction.
Fig.~\ref{fig:sp340} shows also the sensitivity of $\Sigma$ and
$d\sigma/d\Omega$ to the small $E_{1+}$ amplitude for both pion channels. 
The main effects of the $E_{1+}$ multipole in the $p\pi^0$ channel
is found near $\theta_{\pi^0}=90^0$ for the photon asymmetry and near 
$0^0,~90^0$ and $180^0$ for the differential cross section. 
This sensitivity has its origin in the interference term 
${\rm Re}(E_{1+}M_{1+}^*)$, which appears in the coefficients
\begin{eqnarray}
A &\simeq& \frac{5}{2} |M_{1+}|^2 - 3 {\rm Re}(E_{1+}M_{1+}^*)~, \\ 
C &\simeq& - \frac{3}{2} |M_{1+}|^2 + 9 {\rm Re}(E_{1+}M_{1+}^*)~, \\ 
A_{\Sigma} &\simeq& \frac{3}{2} |M_{1+}|^2 + 3 {\rm Re}(E_{1+}M_{1+}^*)~,  
\end{eqnarray}
and contributes to  both $d\sigma/d\Omega$ and $\Sigma$.
Without $E_{1+}$ strength the shape of the differential cross section
would be determined by $C/A = - 3/5$; and the photon asymmetry by 
$\Sigma(90^0) = A_{\Sigma}/A = 3/5$. 
For $E_{1+}/M_{1+} = -2.5\%$ these values would change
to $C/A = -0.7$ and  $\Sigma(90^0) = 0.54$. 
Setting the $E_{1+}$ multipole to zero (dotted line) results in an 
increased differential cross section at
the extreme angles $0^0$ and $180^0$ and a lower value at $90^0$.
The photon asymmetry in contrast is enhanced at $90^0$ for $E_{1+}=0$. 
The same behavior for $\Sigma$ is found in the $n\pi^+$ channel. 
In the differential cross section for $n\pi^+$ the effect of $E_{1+}$ is 
most sizeable at the backward angles.

From the above discussion it becomes obvious that a reliable extraction of 
the small $E_{1+}$ amplitude will require precise photon asymmetry data 
around $90^0$ and precise data for the differential cross section at all 
angles.

\section {\bf Discussion of the different analyses}

We have performed three different analyses of our data in order
to extract the $s$-- and $p$--wave amplitudes and the $E2/M1$ ratio. 
First, a fit to the differential cross 
section and photon asymmetry only in the $\gamma p \rightarrow p \pi^0$
channel; second, an energy independent simultaneous multipole analysis of 
the $\gamma p \rightarrow p \pi^0$  and $\gamma p \rightarrow n \pi^+$ 
data to get the isospin separation 
in the whole energy region ($270-420 {\rm MeV}$).
In the third analysis an energy dependent fixed--t dispersion 
analysis has been performed, 
which includes additional observables in order to study
the stability of the different multipole solutions.

\subsection {\bf Analysis of the MAMI $p\pi^0$ data}

Since the result from the fit to the $\gamma p \rightarrow p\pi^0$ channel 
alone has already been published~\cite{Bec97}, we summarize here only the 
key points of our analysis.
The $p\pi^0$ angular distributions for the unpolarized cross section
$d\sigma_0/d\Omega$, the parallel part $d\sigma_{\parallel}/d\Omega$ 
(pion detected in the plane defined by the photon polarization and 
the photon momentum vector),
and perpendicular part $d\sigma_{\perp}/d\Omega$ can be expressed in the
$s$-- and $p$--wave approximation by the parametrization 
\begin{equation}
\frac{d\sigma_{j}(\theta)}{d\Omega} = \frac{q}{k}(A_j ~+~ B_j cos(\theta) ~+~ 
C_j cos^2(\theta)) \; ,
\end{equation}
where $q$ and $k$ denote the center of mass momenta of the pion and
the photon, respectively, and $j$ indicates the parallel ($\parallel$),
perpendicular ($\bot$) and unpolarized ($0$) components.
The coefficients $A_j$, $B_j$ and $C_j$ are quadratic or bilinear functions
of the $s$-- and $p$--wave amplitudes. In particular,
$d\sigma_{\parallel} / d\Omega$ is sensitive to the $E_{1^+}$ amplitude,
because of interference with $M_{1+}$ in the terms
\begin{eqnarray}
A_{\parallel} &=& \mid E_{0+} \mid^2 + \mid 3E_{1+} - M_{1+} + M_{1-} \mid^2
\; , \\
B_{\parallel} &=& 2{\rm Re}[E_{0+}(3E_{1+} + M_{1+} - M_{1-})^*] \; , \\
C_{\parallel} &=& 12{\rm Re}[E_{1+} (M_{1+} - M_{1-})^*] \; .
\end{eqnarray}
Furthermore, the ratio 
\begin{equation}
R = \frac{1}{12}\frac{C_{\parallel}}{A_{\parallel}} = 
\frac{{\rm Re}(E_{1+}(M_{1+} - M_{1-})^*)}
{\mid E_{0+}\mid^2 + \mid 3E_{1+} +  M_{1+} - M_{1-}\mid^2}
\end{equation}
is very close to the ratio $R_{EM} = E_{1+}^{3/2} / M_{1+}^{3/2}$ as
we will show below.
At the $\Delta(1232)$ resonance ($\delta_{33}=90^0$)
the ratio R can be identified with the $R_{EM}$ value, 
\begin{equation}
R \simeq R_{EM} = \left . \frac{{\rm Im}E_{1+}^{3/2}}{{\rm
Im}M_{1+}^{3/2}}\right|_{W=M_\Delta} \; .
\end{equation}
This is the key point of our analysis.
The method has the advantage of being independent of absolute
normalization and insensitive to many systematic errors, because
$R_{EM}$ is extracted from the ratio of the coefficients $C_{\parallel}$ and
$A_{\parallel}$ fitted to the angular distribution of
$d\sigma_{\parallel}/d\Omega$.
Analyzing the data by this method the following result is
obtained~\cite{Bec97} :
\begin{equation}
R_{EM} = - (2.5 \pm 0.2_{stat}) \% \, .
\end{equation}
Without any correction for neglecting the isospin $1/2$ contributions to
$E_{1+}$ and $M_{1+}$, the maximum absolute systematic error for $R_{EM}$ 
would be $\pm0.5\%$. 
To further reduce this systematic error one can take
the Born contribution and estimate the size of the isospin $1/2$
contribution for $E_{1+}$ and $M_{1+}$. With a conservative estimate the
absolute systematic error reduces to less than $\pm0.2\%$ (see
Appendix B for details).

\subsection {\bf Energy independent multipole analysis of the MAMI $p\pi^0$
and $n \pi^+$ data}

To obtain the isospin decomposition for the $s$ and
$p$ waves in the whole energy range ($270 - 420 {\rm MeV}$), we
have performed an energy independent multipole analysis of our data. 
In this analysis we fit $8$ parameters, the absolute values of the $s$ and 
$p$ waves for isospin $1/2$ and $3/2$
($E_{0+}^{1/2}$, $E_{0+}^{3/2}$,
$M_{1+}^{1/2}$, $M_{1+}^{3/2}$,
$E_{1+}^{1/2}$, $E_{1+}^{3/2}$,
$M_{1-}^{1/2}$ and $M_{1-}^{3/2}$), to the
photon asymmetry and the differential cross section
for the $\gamma p \rightarrow p\pi^0$ and
$\gamma p \rightarrow n\pi^+$ reaction channels.
Higher partial waves ($l_{\pi}\ge2$) are taken into account by
the Born terms, including $\rho$ and $\omega$ exchange in the $t$-channel. 
We obtain the coefficients $A$, $B$ and $C$ 
from the angular distribution of the differential cross section and 
$A_{\Sigma}$ from the photon asymmetry. 
The two pion reaction channels are decribed by $8$ coefficients, which are
independent combinations of the isospin $1/2$ and $3/2$ multipole 
amplitudes of the $s$ and $p$ waves.
In addition the Fermi--Watson theorem is used to determine the real and 
imaginary parts of the $s$-- and $p$--wave multipole amplitudes. 

\begin{figure}
\centerline{\psfig{figure=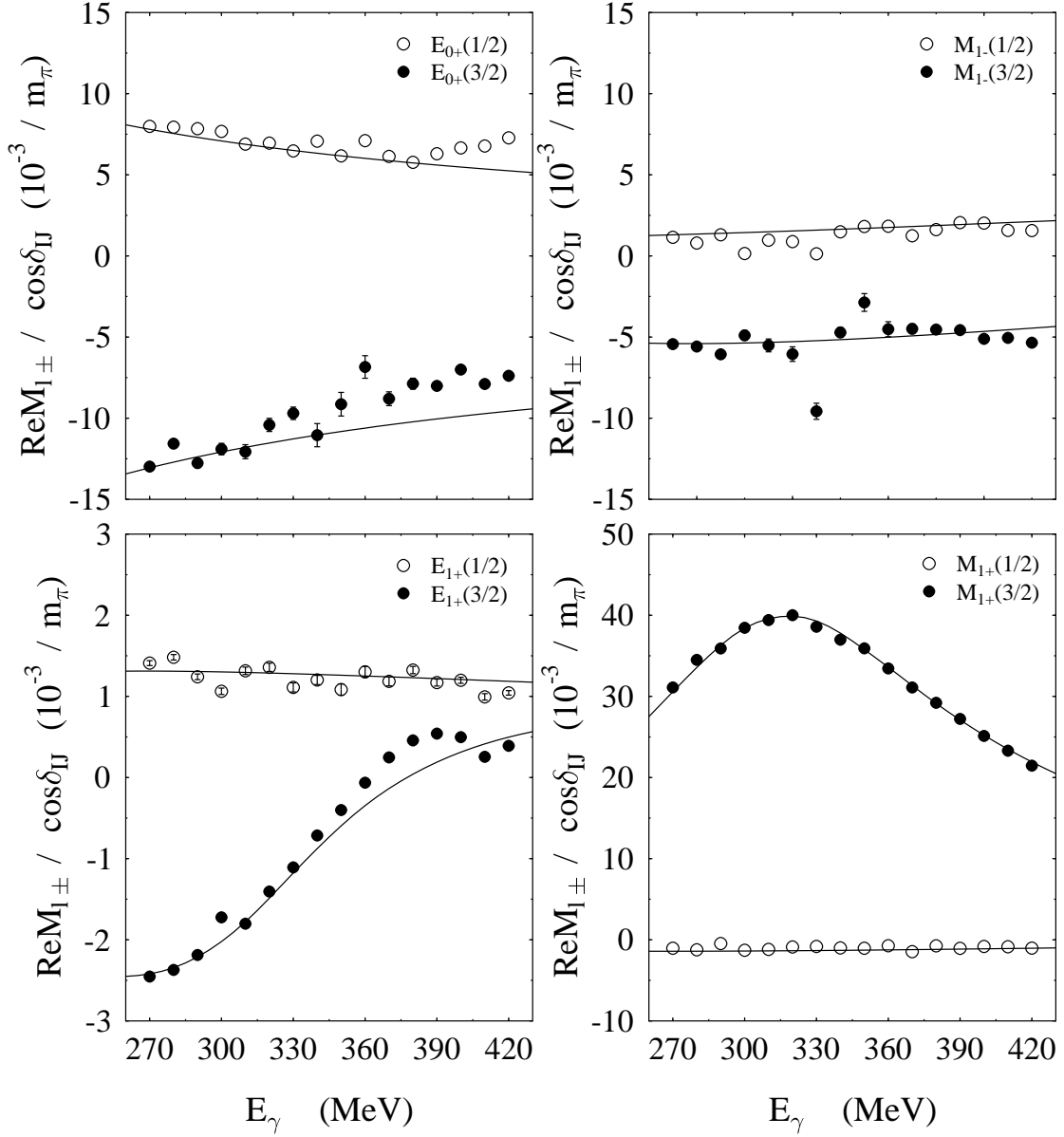,width=15.0cm}}
\caption{The isospin $I=1/2$ and $I=3/2$ components of the $s$-- and $p$--wave
         multipoles for the
         proton\protect\cite{Kra96} in compared to the energy
				 dependent result of the fixed--t dispersion
				 analysis\protect\cite{Han98}. The plotted quantities
				 ${\rm Re}{\cal M}_{l\pm}^I / cos\delta_{l\pm}^I = 
				 {\cal M}_{l\pm}^{I} e^{-i\delta_{l\pm}^I}$
         are equal to the absolute value $\mid {\cal M}_{l\pm}^{I} \mid$ 
				 up to a sign, see eq. (4).}
\label{fig:iso}
\end{figure}

In Fig.~\ref{fig:iso} we show our fitted multipole amplitudes as
${\rm Re}{\cal M}_{l\pm}^I / cos\delta_{l\pm}^I$,
which is the absolute value of the amplitude up to a sign that can change
with photon energy as in the case of
$E_{1+}^{3/2}$. The values of the $s$-- and $p$--wave amplitudes
are fitted independently at each energy to the MAMI data
$\gamma p \rightarrow p \pi^0$ and $\gamma p \rightarrow n \pi^+$.
The results for the isospin
components for the $s$-- and $p$--wave amplitudes are presented
for 16 photon energy bins\cite{Kra96}. Although systematic errors are
greatly reduced because the $p\pi^0$ and $n\pi^+$ data are obtained with the
same set up, this method is sensitive to small remaining
systematic differences in the results of $d\sigma/d\Omega$ and $\Sigma$
for the two pion channels, because it relates 8 parameters to the 8 
``observables'' (A,B,C and $A_{\Sigma}$ 
coefficients for the two pion channels). Including additional polarization
observables in the database and performing an energy dependent multipole
analysis overcomes these shortcomings.

\subsection {\bf Energy dependent multipole analysis of the MAMI data}

In a third analysis the MAMI pion photoproduction data are analyzed
using fixed--t dispersion relations based on Lorentz invariance, isospin
symmetry, unitarity and crossing symmetry~\cite{Han96,Han98}.
This analysis includes the recent MAMI data for the differential
cross section $d\sigma/d\Omega$ and photon asymmetry $\Sigma$ for $p\pi^0$
and $n\pi^+$ from the proton~\cite{Kra96,Fuc96,Hae96}, both older and more
recent data from Bonn for the target asymmetry $T$ 
\cite{Dut96,Men77,Bue94}, and differential cross section data 
on $\pi^-$ production off the neutron
from Frascati~\cite{Car73} and recently from TRIUMF~\cite{Bag88}.

\begin{figure}
\centerline{\psfig{figure=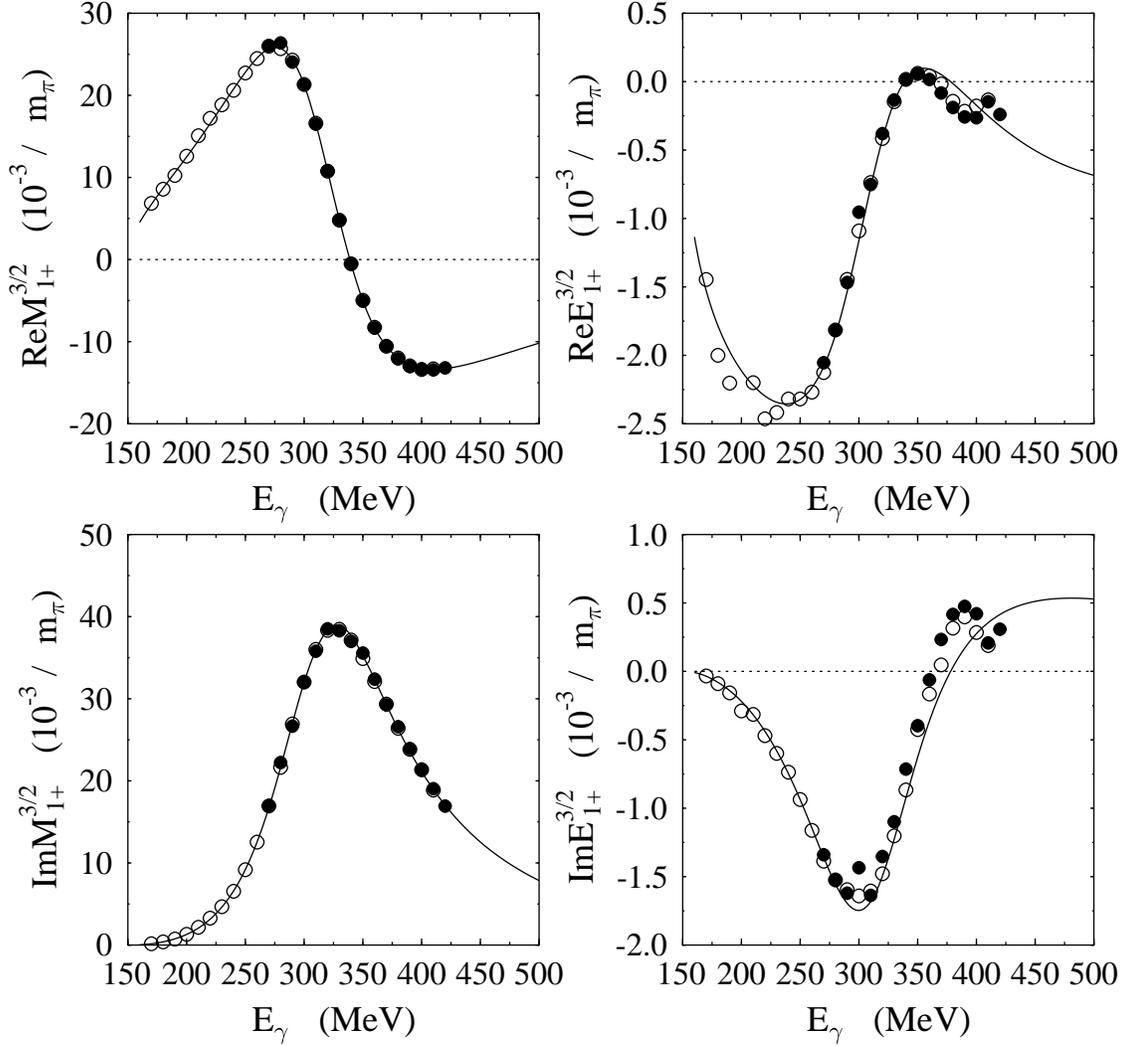,width=15.0cm}}
\caption{The real and imaginary isospin $I=3/2$ components of the $M_{1+}$
         and $E_{1+}$ multipoles.
         The solid dots show the result from the energy independent 
				 fit only to the MAMI data ($d\sigma/d\Omega$ and 
				 $\Sigma$)\protect\cite{Kra96}. The solid line and open circles 
				 show the energy dependent and energy independent
         results from the fixed--t dispersion analysis\protect\cite{Han98}.}
\label{fig:e2m1}
\end{figure}

With this method we performed both an energy dependent and an energy
independent multipole fit as shown
in Fig.~\ref{fig:e2m1} by open circles (energy independent)
and the solid line (energy dependent).
The agreement between the $E_{1^+}^{3/2}$ and
$M_{1^+}^{3/2}$ multipole results obtained from the dispersion analysis and
the multipole analysis described in the previous section is very good.
Both energy independent
results show the same energy behavior for the two multipoles.
For the real and imaginary parts of the $M_{1^+}^{3/2}$ amplitude the
two results are in
excellent agreement, the solid dots lying on top of the open circles up to
$425 {\rm MeV}$ (see
Fig.~\ref{fig:e2m1}). The agreement is also very good in the case of the 
small 
$E_{1^+}^{3/2}$ amplitude. In particular, we find that the inclusion of 
other observables in addition to the MAMI pion
photoproduction data decreases the fluctuations slightly for the 
$E_{1^+}^{3/2}$ multipole.

The results of the energy dependent fit (solid line in
Fig.~\ref{fig:e2m1}) agree well with the $M_{1+}$ multipole derived from
the energy independent fit, but small systematic differences may be seen 
for the electric multipole $E_{1+}$ at energies
above the $\Delta(1232)$ resonance ($E_{\gamma} \ge 360 {\rm MeV}$).
The ratio $R_{EM}$ at the resonance agrees quite well in both dispersion 
analyses,
\begin{equation}
R_{EM} = - (2.5 \pm 0.1_{stat}) \%
\end{equation}
for the energy dependent fit and
\begin{equation}
R_{EM} = - (2.33 \pm 0.17_{stat}) \%
\end{equation}
for the energy independent fit.

In addition we have checked the effect of changes in the database. 
Replacing our differential cross section by the Bonn data obtained in 
the  seventies
\cite{Fis72,Gen74} reproduces the results for the leading
multipoles but changes the smaller multipoles to some degree (see
Tab.~\ref{tab:rem}).
The Bonn results for $d\sigma/d\Omega$ at very forward and backward angles
force the fit to a smaller $E_{1+}$ value near the resonance.
However, it should be kept in mind that the overall compilation of the Bonn
$\gamma p \rightarrow p \pi^0$ data by Genzel et al. \cite{Gen74} results
from different experimental setups, which have been re--appraised and 
re--evaluated before combining them into one data set. The main part of the 
angular distribution ($50^0\le \theta_{\pi} \le160^0$) for 
$d\sigma/d\Omega$ was measured
by detecting the recoil proton in a magnetic spectrometer \cite{Fis71}, i.e.
by fixed angle and single energy measurements. 
The differential cross
section for $10^0\le \theta_{\pi} \le 70^0$ was measured by detecting the
two decay photons of the $\pi^0$ with lead glass blocks in a different
experiment \cite{Hil74}. Each of these experiments has a systematic
error of the order of $\pm 5\%$, and it takes much less than this systematic 
error to change the shape of $d\sigma/d\Omega$ ($C/A$ ratio)
and with this the $R_{EM}$ value.

In the case of the LEGS data the larger cross section leads
to larger values for the resonant multipoles $M_{1+}^{3/2}$ and
$E_{1+}^{3/2}$.
The main reason for this is that the cross section difference between the LEGS
and MAMI data is energy dependent (resonance behavior) as shown in
Fig.~\ref{fig:pa90} and Fig.~\ref{fig:pa85}.
Inclusion of the polarization data $T$, $\Sigma$ and $P$ from 
Kharkov \cite{Kha80} does not affect our fit 
because of the large statistical and systematic errors especially for the 
$\gamma p \rightarrow p \pi^0$ channel.

\section {\bf Discussion of the results}

\begin{figure}
\centerline{\psfig{figure=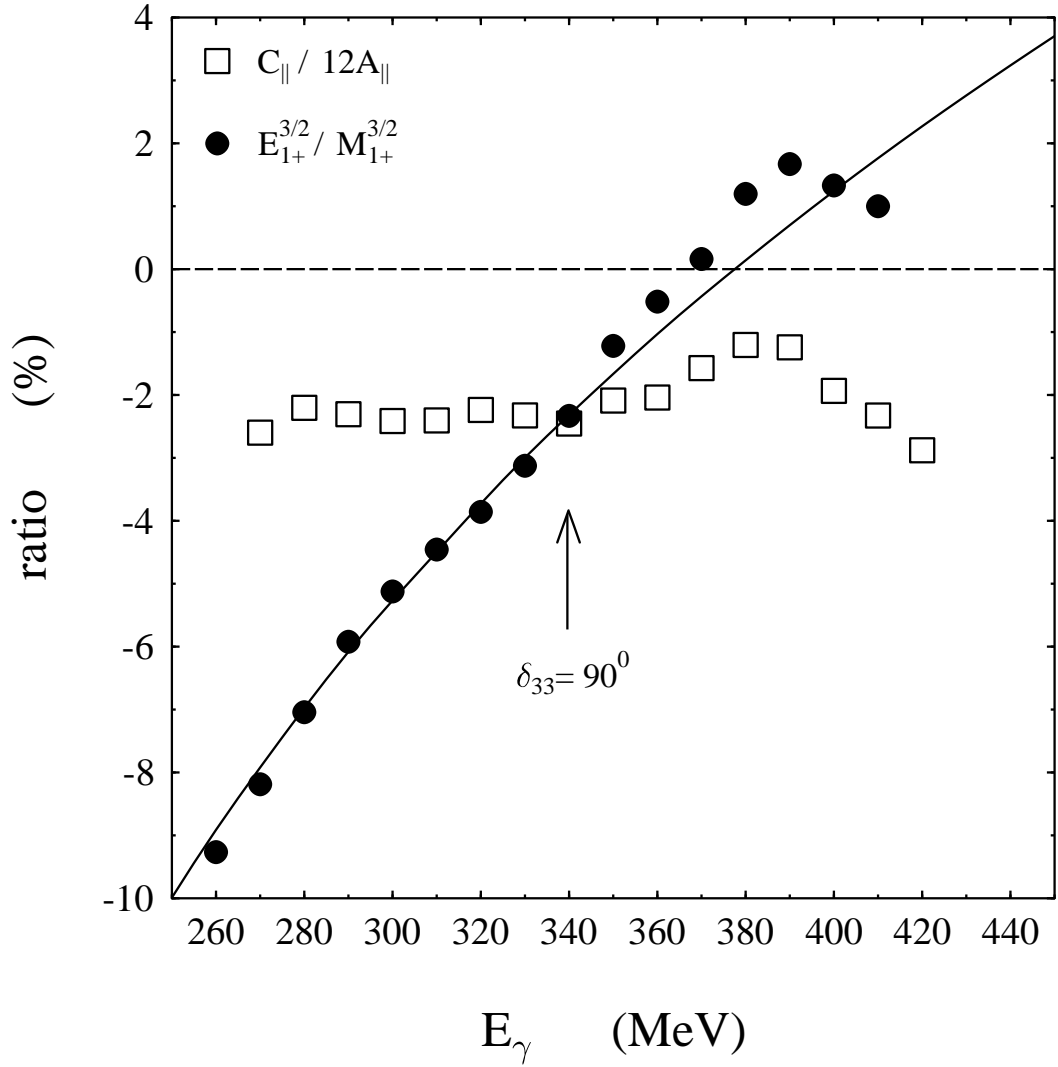,width=15.0cm}}
\caption{The energy dependence of the ratio $E_{1+}^{3/2} / M_{1+}^{3/2}$.
         The solid dots (energy independent) and the solid line 
				 (energy dependent)
         present the result from the fixed--t dispersion multipole analysis
         \protect\cite{Han98}. 
         In addition the energy dependence of $R = C_{||}/(12 A_{||})$ 
				 is shown as open squares.}
\label{fig:e2m1_pa}
\end{figure}

According to the Fermi--Watson theorem the $E_{1+}^{3/2}$ and 
$M_{1+}^{3/2}$ partial waves have the same phase $\delta_{33}$
and the ratio $E_{1+}^{3/2} / M_{1+}^{3/2}$
is a real quantity. As shown in Fig.~\ref{fig:e2m1_pa},
this ratio is strongly dependent on the photon energy and varies
from $-8\%$ at $E_{\gamma} = 270 {\rm MeV}$ to 
$+2\%$ at $E_{\gamma} = 420 {\rm MeV}$.
The ratio $R_{EM}$ is defined at the $\Delta(1232)$ resonance 
position, where $\delta_{33} = 90^0$, by 
\begin{equation}
R_{EM} = \left. \frac{E_{1+}^{3/2}}{M_{1+}^{3/2}}\right|_{W=M_\Delta} = 
\left. \frac{{\rm Im}E_{1+}^{3/2}}{{\rm Im}M_{1+}^{3/2}}\right|_{W=M_\Delta} 
\, .
\end{equation}
We note that within the K--matrix fomalism this ratio is free of background
contributions since $M_\Delta$ is the K--matrix pole.
The extraction of the genuine $\Delta(1232)$--resonance parts of
the magnetic dipole and electric quadrupole multipoles is model dependent, see
for example the work on dynamical models~\cite{Ber93,Sat96,Wil96} and the 
recently proposed speed plot analysis~\cite{Han96,Wil98}.
In order to get a reliable value for $R_{EM}$ it is important
to use a database with small angle dependent systematic errors,
because the small $E_{1+}$ amplitude depends on the shape of
$d\sigma/d\Omega$ and the absolute magnitude of the photon asymmetry at 
$90^0$.
To obtain a consistent separation for the isospin $1/2$ and $3/2$ parts of
the partial waves, experimental data 
are required with small inconsistencies in the absolute normalization between
the $\gamma p \rightarrow p \pi^0$ and $\gamma p \rightarrow n \pi^+$
observables. To reduce the influence of such systematics in the
multipole analysis, the observables $\Sigma$ and $d\sigma/d\Omega$ for 
both reaction channels $p \pi^0$ and $n \pi^+$ were simultaneously measured 
at all angles and energies with the  DAPHNE detector at MAMI.
The three different analyses to extract $R_{EM}$ from 
the MAMI data agree very well with each other, 
and at the $\Delta(1232)$ resonance we get the final result 
\begin{equation}
R_{EM} =  
\frac{{\rm Im}E_{1+}^{3/2}}{{\rm Im}M_{1+}^{3/2}} = 
- (2.5 \pm 0.1_{stat} \pm 0.2_{sys})\% \, . 
\end{equation}
In addition, the discussions about the extraction of the correct
$R_{EM}$ ratio from our data, which arose in the literature after our first
publication~\cite{Bec97} can now be summarized as follows.
In the Comment of the VPI group~\cite{Wor97}, our $p \pi^0$ data were
included in the SAID database and a value $R_{EM} = - (1.5 \pm 0.5) \%$ was
obtained. As pointed out in our Reply~\cite{Bec97_1} the difference
between the VPI result and our value $R_{EM}=-(2.5 \pm 0.2)\%$ is due to
the database used in the SAID analysis. As has been recently confirmed 
by the VPI group~\cite{Wor97_1}, the exclusion of all pre--$1980$ 
$p\pi^0$ differential cross section data in the SAID database changes 
$R_{EM}$ to $-2.5\%$. It is obvious that the key problem arises from the 
inconsistencies in the different data sets used in the analysis.
The forward and backward angles for the differential cross sections used in
the SAID database force the VPI solution to smaller $R_{EM}$ values. 
However, this solution overpredicts the Mainz photon asymmetry data at 
resonance,
$\Sigma(90^0) = 0.58$, compared to the Mainz result, $\Sigma(90^0) = 0.54$.
The reason is that at resonance ($\delta_{33}=90^0$) the fit has only one
parameter, the $R_{EM}$ value, to describe simultaneously the shape of the
differential cross section 
\begin{equation}
\frac{C}{A} \simeq \frac{-3/2 + 9 R_{EM}}{5/2 - 3 R_{EM}}  
\end{equation}
and the photon asymmetry at $90^0$
\begin{equation}
\frac{A_{\Sigma}}{A} \simeq \frac{3/2 + 3 R_{EM}}{5/2 - 3 R_{EM}} \, .  
\end{equation}
As we have stated before, the compilation of all the existing 
experimental data into one database can result in
mean values for the dominant multipoles and meaningless values
for the small multipoles. For example, if one combines the new MAMI
data with the Bonn differential cross section results, then the 
$R_{EM}$ value is affected by small systematic differences in the two 
data sets. The Bonn differential cross sections range from 
$10^0$ to $160^0$ while the MAMI cross sections cover angles from
$75^0$ to $125^0$ and are slightly below the Bonn data at
$E_{\gamma} = 340 {\rm MeV}$.
In the combined data set the shape of the differential cross section 
is changed, i.e. the $C/A$ ratio gets smaller. 
This is the main reason that for the combined
Bonn and MAMI data set the $R_{EM}$ value is even below the value which 
one gets from the Bonn data alone.
This influence of the Bonn data on the extracted value of the ratio $R_{EM}$
has been confirmed by the BNL\cite{Bla97} and RPI/VPI groups\cite{Dav98}.

In a second Comment~\cite{Dav97} our $p \pi^0$ data were analyzed by the RPI
group, who obtained the result $R_{EM}=-(3.2 \pm 0.25)\%$. However, the
inclusion of our $n \pi^+$ data in the database lowered this value to
$R_{EM}=-(2.64 \pm 0.25)\%$~\cite{Dav97_1}, in agreement with our analysis.

\section {\bf Summary and Outlook}

We made the first simultaneous and accurate measurements of
the differential cross sections and photon asymmetries
for the reactions $p(\vec{\gamma},p)\pi^0$ and
$p(\vec{\gamma},\pi^+)n$ in the $\Delta(1232)$ region. 
Using this data we have performed a multipole analysis to obtain the
isospin decomposition of the $s$-- and $p$--wave multipoles and extract the 
$E2/M1$ ratio over the energy range $270-420{\rm MeV}$.
Our final results at the $\Delta(1232)$ resonance position 
($\delta_{33} = 90^0$)
are $R_{EM} = - (2.5\pm0.1_{stat}\pm0.2_{sys})\%$, and 
$A_{1/2} = -(131 \pm 1)(10^{-3}/\sqrt{GeV})$ and
$A_{3/2} = -(251 \pm 1)(10^{-3}/\sqrt{GeV})$ for the helicity amplitudes.

In the meantime more $p(\vec{\gamma},\pi^0)p$ data~\cite{Leu99} for the 
differential cross section and the photon asymmetry have been taken. This
new experiment covers the full range of polar angles by observing the two 
decay photons in the TAPS detector at MAMI. The analysis is in 
progress and will produce new differential cross sections for the extreme 
forward and backward angles. In this way we hope also to clarify the absolute 
normalization problem in the differential cross section between LEGS and 
MAMI.

{\bf Acknowledgments}

The authors wish to acknowledge the excellent support of K.H. Kaiser
and H. Euteneuer and the accelerator group of MAMI, as well as many 
other scientists and technicians of the Institut f\"{u}r Kernphysik 
at Mainz, DAPNIA/ SPhN at Saclay and INFN at Pavia.
We would also like to thank   
H. Arenh\"ovel, H. Schmieden and P. Wilhelm
for fruitful discussions and comments.
This work was supported by the Deutsche Forschungsgemeinschaft (SFB~201) and
the UK Engineering and Physical Sciences Research Council.

\newpage
{\bf Appendix A}

The coefficients $A$, $B$, $C$ and $A_\Sigma$ in the $s$-- and $p$--wave 
approximation for the differential cross section and the photon asymmetry:
\begin{eqnarray*}
A(s_{wave},p_{wave}) & = & |E_{0+}|^2 + \frac{9}{2} |E_{1+}|^2 + \frac{5}{2} 
|M_{1+}|^2 - |M_{1-}|^2
- 3{\rm Re}(E_{1+}(M_{1+} - M_{1-})^*) + {\rm Re}(M_{1-} M_{1+}^*) \\ 
B(s_{wave},p_{wave}) & = & 2{\rm Re}(E_{0+}(3E_{1+} + M_{1+} - M_{1-})^*) \\  
C(s_{wave},p_{wave}) & = & \frac{9}{2} |E_{1+}|^2 - \frac{3}{2} |M_{1+}|^2  
+ 9{\rm Re}(E_{1+}(M_{1+} - M_{1-})^*) - 3{\rm Re}(M_{1-} M_{1+}^*) \\  
A_\Sigma(s_{wave},p_{wave}) & = & - \frac{9}{2} |E_{1+}|^2 + \frac{3}{2} 
|M_{1+}|^2 + 3{\rm Re}(E_{1+}(M_{1+} - M_{1-})^*) + 
3{\rm Re}(M_{1-} M_{1+}^*) \\ 
\end{eqnarray*}

The differential cross section up to $l_{\pi}=2$ partial waves 
\begin{eqnarray*}
\frac{d\sigma}{d\Omega} = \frac{q}{k}( A + B cos(\theta) + C cos^2(\theta) +
D cos^3(\theta) + E cos^4(\theta))  
\end{eqnarray*}
has five coefficients with

\begin{eqnarray*}
A & = & A(s_{wave},p_{wave}) \\ 
  &   & + \frac{5}{2}|E_{2-}|^2 + \frac{45}{4}|E_{2+}|^2 
        + \frac{9}{2} |M_{2-}|^2 + \frac{9}{2} |M_{2+}|^2 
        + {\rm Re}(E_{0+}(-6E_{2+} - 3M_{2+} - E_{2-} + 3M_{2-})^*) \\   
  &   & + {\rm Re}(E_{2-}(\frac{15}{2}E_{2+} + 3M_{2-} + 3M_{2+})^*) 
        + {\rm Re}(M_{2-}(-9M_{2+} - \frac{9}{2}E_{2+})^*)
        + {\rm Re}(\frac{9}{2}M_{2+}E_{2+}^*) \\
B & = & B(s_{wave},p_{wave}) \\ 
  &   & + 2{\rm Re}(M_{1+}(E_{2-} - 9E_{2+} + 9M_{2+} + 6M_{2-})^*) 
        + {\rm Re}(M_{1-}(18E_{2+} - 2E_{2-} + 9M_{2+} + 6M_{2-})^*)\\
  &   & + 6{\rm Re}(E_{1+}(-2E_{2-} + 3M_{2-} + 3M_{2+} + 3E_{2+})^*)\\  
C & = & C(s_{wave},p_{wave}) \\
  &   & - \frac{3}{2} |E_{2-}|^2 + \frac{27}{2} |E_{2+}|^2  
        + \frac{9}{2} |M_{2-}|^2 + 27 |M_{2+}|^2 
        + {\rm Re}(E_{0+}(6E_{2+} + 3M_{2+} + E_{2-} - 3M_{2-})^*) \\  
  &   & + {\rm Re}(E_{2-}(-63E_{2+} - 9M_{2-} + 9M_{2+})^*) 
        + {\rm Re}(M_{2-}(81E_{2+} + 81M_{2+})^*)
        - {\rm Re}(81M_{2+}E_{2+}^*) \\
D & = & 2{\rm Re}(M_{1+}(30E_{2+} - 12M_{2+} - 18M_{2-})^*) 
        + {\rm Re}(M_{1-}(-30E_{2+} - 15M_{2+})^*) \\
  &   & + {\rm Re}(E_{1+}(-36M_{2-} + 36M_{2+} + 18E_{2+} + 18E_{2-})^*) \\  
E & = & \frac{45}{4} |E_{2+}|^2 - \frac{45}{2} |M_{2+}|^2  
        - 90{\rm Re}(M_{2+} M_{2-}^*)  
        + \frac{45}{2}{\rm Re}(E_{2+} (5M_{2+} - 5M_{2-} + 3E_{2-})^*) \\
\end{eqnarray*}

The photon asymmetry $\Sigma$
\begin{eqnarray*}
\Sigma \frac{d\sigma}{d\Omega} \frac{1}{sin^2\theta} = \frac{q}{k}
(A_{\Sigma} + B_{\Sigma} cos\theta + C_{\Sigma} cos^2\theta)
\end{eqnarray*}

has three coefficients with

\begin{eqnarray*}
A_{\Sigma} & = & A_{\Sigma}(s_{wave},p_{wave}) \\ 
  &   & - \frac{3}{2} |E_{2-}|^2 - 9 |E_{2+}|^2 - \frac{9}{2} |M_{2-}|^2  
        + {\rm Re}(E_{0+}(E_{2-} + E_{2+} + M_{2-})^*) \\  
  &   & + {\rm Re}(E_{2-}(3M_{2-} + \frac{21}{2}E_{2+})^*) 
        - \frac{9}{2} {\rm Re}(E_{2+} M_{2-}^*) \\ 
B_{\Sigma} & = & + 3{\rm Re}(M_{1+}(5E_{2+} + 4M_{2+} + 6M_{2-})^*) 
        - 3{\rm Re}(M_{1-}(5E_{2+} - 5M_{2+})^*)\\ 
  &		& - 3{\rm Re}(E_{1+}(9E_{2+} - 9E_{2-})^*) \\  
C_{\Sigma} & = & \frac{45}{2} |M_{2+}|^2 - 45 |E_{2+}|^2  
        + \frac{45}{2} {\rm Re}(E_{2+}(3E_{2-} - 2M_{2-} - M_{2+})^*)  
        + 90{\rm Re}(M_{2+} 3M_{2-}^*) \\ 
\end{eqnarray*}

\newpage
{\bf Appendix B}

At the $\Delta(1232)$ resonance position, where the phase $\delta_{33}$
passes through $90^0$ ($E_{\gamma} \simeq 340{\rm MeV}$) we find
${\rm Re}M_{1^+}(3/2) = 0$, ${\rm Re}(M_{1^+} - M_{1^-}) \simeq 0$ and 
negligible contributions from higher partial waves ($l_\pi \ge 2$) 
(see sect. IV b).
The ratio $R$ of the coefficients $A_{||}$ and $C_{||}$
for $p(\vec{\gamma},p)\pi^0$ can then be expressed by
\begin{eqnarray*}
R = \frac{C_{\|}} {12A_{\|}} = \frac{{\rm Re}(E_{1+}(M_{1+} - M_{1-})^*)}
{|E_{0+}|^2 + |3E_{1+} - M_{1+} + M_{1-}|^2} \; . 
\end{eqnarray*}
Neglecting $|E_{0+}|^2$, $9|E_{1+}|^2$ and all terms with ${\rm Re}(M_{1+} -
M_{1-})$ results in
\begin{eqnarray*}
R = \frac{{\rm Im}E_{1+}{\rm Im}(M_{1+} - M_{1-})}
{-6{\rm Im}E_{1+}{\rm Im}(M_{1+} - M_{1-}) + {\rm Im}^2(M_{1+} - M_{1-})} \; . 
\end{eqnarray*}
Dividing by ${\rm Im}^2(M_{1+} -M_{1-})$ we obtain
\begin{eqnarray*}
R = \frac{R_{\pi^0}}{1 - 6R_{\pi^0}}  , \; 
\end{eqnarray*}
with
\begin{eqnarray*}
R_{\pi^0} = \frac{{\rm Im}E_{1+}} {{\rm Im}M_{1+} - {\rm Im}M_{1-}} \; . 
\end{eqnarray*}
If we further neglect ${\rm Im}M_{1-}$ and the isospin $1/2$ components
${\rm Im}M_{1+}^{1/2}$ and ${\rm Im}E_{1+}^{1/2}$, then
the largest correction comes from ${\rm Im}E_{1+}^{1/2}$,
which is of the order of
$10-20\%$ of ${\rm Im}E_{1+}$.
The final result depends on the ratio of
${\rm Im}E_{1+}^{1/2} / {\rm Im}E_{1+}(p\pi^0)$. If we further assume 
$R = -2.5 \%$ we obtain:

(a) for ${\rm Im}E_{1+}^{1/2} = 0$ 
\begin{eqnarray*}
R = \frac{C_{\|}} {12A_{\|}} \simeq \frac{1.R_{EM}}{1 - 6.R_{EM}} 
~\rightarrow~ R_{EM} = -2.95\% , \; 
\end{eqnarray*}

(b) for ${\rm Im}E_{1+}^{1/2} / {\rm Im}E_{1+} ~=~ 10\%$  
\begin{eqnarray*}
R = \frac{C_{\|}} {12A_{\|}} \simeq \frac{1.1R_{EM}}{1 - 6.6R_{EM}} 
~\rightarrow~ R_{EM} = -2.65\%  , \; 
\end{eqnarray*}

(c) for ${\rm Im}E_{1+}^{1/2} / {\rm Im}E_{1+} ~=~ 20\%$
\begin{eqnarray*}
R = \frac{C_{\|}} {12A_{\|}} \simeq \frac{1.2R_{EM}}{1 - 7.2R_{EM}} 
~\rightarrow~ R_{EM} = -2.45\% , \; 
\end{eqnarray*}
with
\begin{eqnarray*}
R_{EM} = \frac{ {\rm Im}E_{1+}^{3/2} } { {\rm Im}M_{1+}^{3/2} } \; . 
\end{eqnarray*}
Without any correction for the isospin $1/2$ contributions to
$E_{1+}$ and $M_{1+}$ the maximum systematic error of $R_{EM}$ would be
$\pm0.5\%$ absolute. To further reduce this systematic error we can estimate 
the size of the isospin $1/2$ contribution by the Born contribution. 
A conservative estimate of the remaining
systematic error is less than $\pm0.2\%$ absolute.

\begin{center}
\begin{table}
\begin{tabular}{cccc}
Data & ${\rm Im}M_{1+}^{3/2}$ & ${\rm Im}E_{1+}^{3/2}$ & $R_{EM}(\%)$ \\ \hline
MAMI & $37.66$ & $-0.924$ & $-2.54$ \\
LEGS & $39.35$ & $-1.276$ & $-3.24$ \\
Bonn & $38.31$ & $-0.643$ & $-1.68$ \\ 
\end{tabular}
\caption{Imaginary parts of the multipoles $E_{1+}^{3/2}$ and $M_{1+}^{3/2}$
in units of $10^{-3} /m_{\pi^+}$, and the ratio $R_{EM} = {\rm Im}E_{1+}^{3/2} / 
{\rm Im}M_{1+}^{3/2}|_{W=M_\Delta}$ in the energy dependent dispersion 
analysis. The different rows correspond to differential cross section data 
from MAMI~\protect\cite{Kra96,Hae96}, 
LEGS~\protect\cite{Bla97} and Bonn~\protect\cite{Fis72,Gen74}.}
\label{tab:rem}
\end{table}
\end{center}
\input{psfig}
%
%Orginal 18cm 25cm
%
%
%Orginal 18cm 25cm
%
%
%Orginal 18cm 25cm
%
%
%Orginal 18cm 25cm
%
%
%Original 
%
%
%Original
%
%
%Orginal 24cm 18cm
%
%
%Orginal 16cm 16cm
%
\begin{figure}
\centerline{\psfig{figure=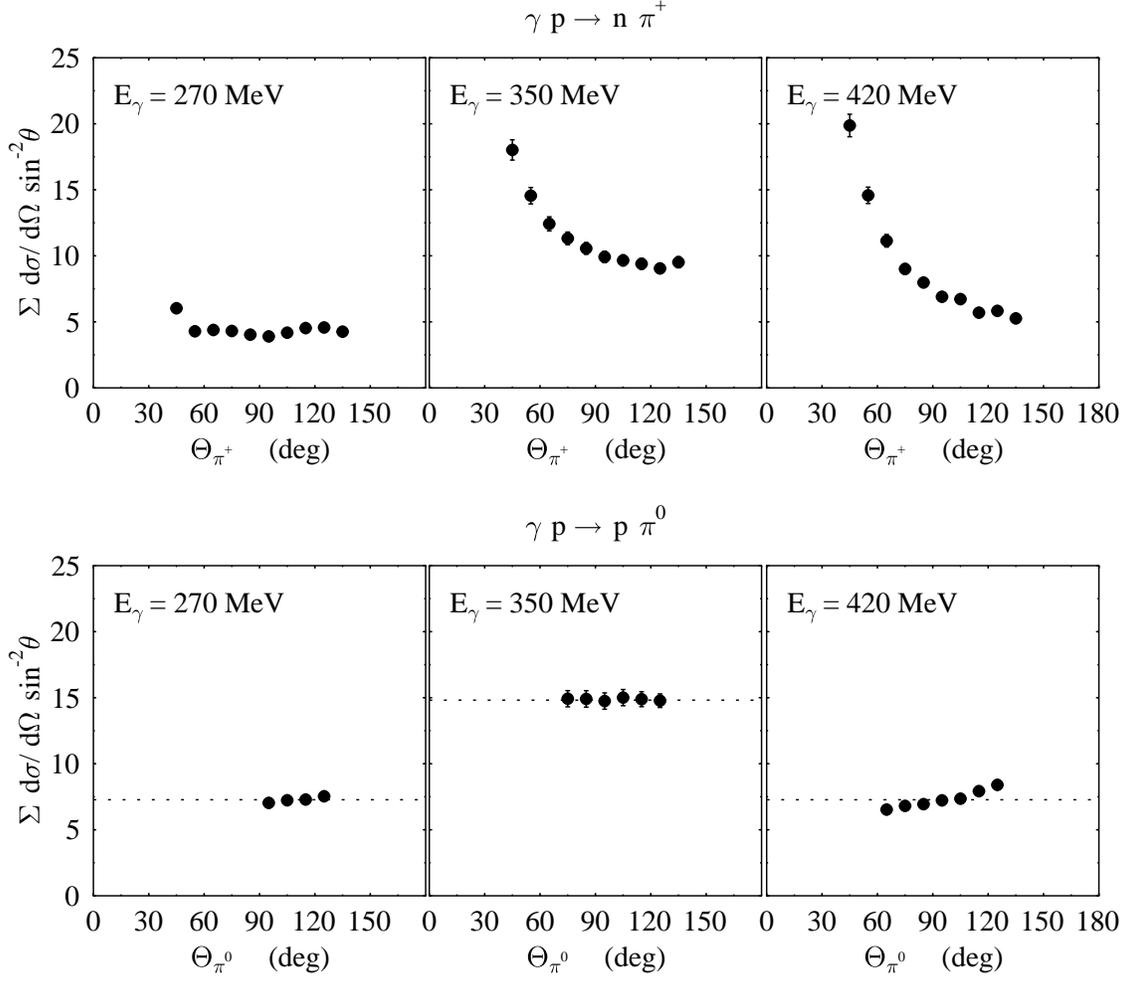,width=15.0cm}}
\caption{The difference of the linear polarization cross sections,
         $(d\sigma_{\bot}/d\Omega - d\sigma_{||}/d\Omega) / (2 sin^2(\theta))$
         $= \Sigma d\sigma / d\Omega sin^{-2}\theta$ for 
         $p(\vec{\gamma},p)\pi^0$ and $p(\vec{\gamma},\pi^+)n$.}
\label{fig:sp}
\end{figure}
%
%Orginal 18cm 9cm
%
\newpage
%
%Original 18cm 24cm
%
%
%
%Original 18cm 24cm
%
%
%Original 11cm 11cm
%
\end{document}